\documentclass[useAMS,usenatbib]{mnras}
\usepackage[fleqn]{amsmath}
\usepackage{amssymb}
\usepackage{wasysym}
\usepackage{graphicx}
\usepackage{grffile}
\usepackage{times}
\usepackage[percent]{overpic}
\usepackage{color}
\usepackage{newtxtext}
\usepackage[T1]{fontenc}
\usepackage{ae,aecompl}
\usepackage{scalerel,stackengine}
\stackMath
\renewcommand* {\vec}[1]{\boldsymbol{#1}} 
\newcommand*{\eps}{\varepsilon}

\renewcommand*{\d} {\mathrm{d}}
\newcommand*{\pot} {\psi} 
\newcommand*{\sub}[2] {{#1}_{\mathrm{#2}}}
\newcommand{\beqn}{\begin{equation}}
\newcommand{\neqn}{\end{equation}} \newcommand{\mathi}{{\,\mathrm{i}}}
\newcommand{\mathe}{{\,\mathrm{e}}} \arraycolsep1pt
\newcommand\reallywidehat[1]{%
\savestack{\tmpbox}{\stretchto{%
  \scaleto{%
    \scalerel*[\widthof{1.2\ensuremath{#1}}]{\kern.1pt\mathchar"0362\kern.1pt}%
    {\rule{0ex}{\textheight}}
  }{\textheight}%
}{2.4ex}}%
\stackon[-5.9pt]{#1}{\tmpbox}%
}

\title[Softening and spiral modes] {How gravitational softening
  affects galaxy stability \\ I. Linear mode analysis of disc
  galaxies}

\author[S. De Rijcke et al.]{Sven De Rijcke$^{1}$\thanks{E-mail:
    sven.derijcke@Ugent.be}, Jean-Baptiste Fouvry$^{2}$\thanks{Hubble
    fellow}, Walter Dehnen$^{3,4}$\\ $^{1}$Ghent University,
  Dept. Physics \& Astronomy, Krijgslaan 281, S9, B-9000, Ghent,
  Belgium\\ $^{2}$Institute for Advanced Study, Einstein Drive,
  Princeton, NJ 08540, USA\\ $^{3}$Department of Physics and
  Astronomy, University of Leicester, Leicester LE1 7RH,
  UK\\ $^{4}$Universit\"ats-Sternwarte der
  Ludwig-Maximilians-Universit\"at, Scheinerstrasse 1, M\"unchen
  D-81679, Germany}

\begin{document}

\def \aj {Astron. J.}  \def \mnras {Mon. Not. R. Astron. Soc.}  \def
\apj {Astrophys. J.}  \def \apjs {Astrophys. J. Suppl.}  \def \apjl
     {Astrophys. J. Let.}  \def \aap {Astron. \& Astrophys.}  \def
     \aaps {Astron. \& Astrophys. Suppl. Ser.}  \def \nat {Nature}
     \def \pasp {Pub. Astron. Soc. Pac.}
\def \araa {ARA\&A}
\date{} \pagerange{\pageref{firstpage}--\pageref{lastpage}}
\pubyear{2009}

\maketitle

\label{firstpage}

\begin{abstract}
Linear perturbation is used to investigate the effect of gravitational
softening on the retrieved two-armed spiral eigenmodes of razor-thin
stellar discs.

We explore four softening kernels with different degrees of gravity
bias, and with/without compact support (compact in the sense that they
yield exactly Newtonian forces outside the softening kernel). These
kernels are applied to two disc galaxy models with well-known
unsoftened unstable modes.  We illustrate quantitatively the
importance of a vanishing linear gravity bias to yield accurate
frequency estimates of the unstable modes. As such, Plummer softening,
while very popular amongst simulators, performs poorly in our tests.

The best results, with excellent agreement between the softened and
unsoftened mode properties, are obtained with softening kernels that
have a reduced gravity bias, obtained by compensating for the
sub-Newtonian forces at small interparticle distances with slightly
super-Newtonian forces at radii near the softening length. We present
examples of such kernels that, moreover, are analytically simple and
computationally cheap. Finally, these results light the way to the
construction of softening methods with even smaller gravity bias,
although at the price of increasingly complex kernels.

\end{abstract}


\section{Introduction}\label{intro}

The evolution of a collisionless stellar system is determined by the
collisionless Boltzmann equation (CBE)
\begin{equation}
	\frac{\partial F(\vec{x},\vec{v},t)}{\partial t} +
        \vec{v}\cdot\frac{\partial F(\vec{x},\vec{v},t)}{\partial
          \vec{x}}+ \frac{\partial V(\vec{x},t)}{\partial
          \vec{x}}\cdot\frac{\partial F(\vec{x},\vec{v},t)}{\partial
          \vec{v}}=0.
\end{equation}
Here, $F(\vec{x},\vec{v},t)$ is the distribution function (DF), which
gives the stellar phase-space density at location $(\vec{x},\vec{v})$
and time $t$. The motion of the collisionless fluid is controlled by
the (positive) binding potential $V(\vec{x},t)$.  The direct numerical
integration of the CBE in six-dimensional phase space is in general
impossible because under the CBE the DF develops ever finer structures
owing to phase mixing or chaotic mixing. However, numerical schemes
which smooth out such fine structure (whereby violating the CBE) are
possible but taxing
\citep{2013ApJ...762..116Y,2014MNRAS.442.3073S,2015MNRAS.450.3724C}.

A much more popular method for modelling collisionless stellar
dynamics is an $N$-body simulation:~a Monte-Carlo approach, which
  integrates the CBE via the method of characteristics. The DF is
  represented by a collection of phase-space points, called
``particles'', and each particle is evolved through phase space along
its characteristic curve by solving the first-order differential
equations
\begin{subequations}
\begin{align}
	\frac{\d \vec{x}}{\d t} &= \vec{v}, \\ \frac{\d \vec{v}}{\d t}
        &= \frac{\partial V(\vec{x},t)}{\partial \vec{x}}.
\end{align}
\end{subequations}
In the case of gravitational forces, the binding potential can be
written as
\begin{subequations}
	\label{eqs:V}
\begin{align}
	V(\vec{x},t) &= G \int \frac{\rho(\vec{x}',t)\,\d
          \vec{x}'}{|\vec{x}-\vec{x}'|} \\ \label{eq:V:unsoft}
        &\approx G \sum_i \frac{m_i}{|\vec{x}-\vec{x}_i|}
\end{align}
\end{subequations}
with $\rho$ the stellar mass density and $m_i$ and $\vec{x}_i$ the
mass and position, respectively, of the $i^{\text{th}}$
particle. Using the approximation~{\eqref{eq:V:unsoft}} to the
gravitational field, diverging accelerations may occur in close
encounters (`collisions') between particles. Such collisions are an
artefact of the much smaller $N$ in the simulation than in the
simulated system. This problem is generally solved by ``softening''
gravity, when the $1/r$ potential in equations~\eqref{eqs:V} is
replaced by a non-diverging form,
\begin{equation}
	\label{eq:soft:kern}
	\frac{1}{r} \;\to\; \pot(r) = \frac{1}{\eps}
        \phi\left(\frac{r}{\eps}\right),
\end{equation}
where $\pot(r)$ is the softening Green's function, $\eps$ the
\emph{softening length}, and $\phi$ the dimensionless \emph{softening
  kernel}. For suitable functions $\phi$ this modifies the
  inter-particle interactions such that accelerations remain bounded
  and strongly deflecting encounters are avoided.

Unfortunately, this force modification results in a \emph{bias} of
gravity and hence changes the character of the physical problem being
addressed by the $N$-body simulation. In practice, a balance must be
found between too much softening, causing force bias, and too little
softening, allowing strong encounters that render the $N$-body
dynamics intractable and collisional
\citep{1996AJ....111.2462M}. \cite{2001MNRAS.324..273D} has derived
asymptotic relations in the context of spherically symmetric systems,
which can be used to inform the choice of the softening parameters
(kernel and length) such as to minimise the resulting mean-square
gravity error.

However, it remains unclear what the optimal choice of these
parameters is in terms of accurately modelling the dynamics, rather
than merely minimising the gravity error. The goal of this series of
papers is to investigate this question by considering stellar
dynamical problems which invoke non-trivial dynamics but are still
simple enough that accurate solutions of the CBE are
available. Specifically, we consider unstable two- and
three-dimensional systems, whose eigenmodes can be accurately obtained
from linear perturbation theory. In this first paper, we focus on the
two-armed (multiplicity $m=2$) spiral-shaped eigenmodes of
two-dimensional razor-thin disc galaxies in the limit of $N\to\infty$
(which can be achieved with computationally cheap linear mode analysis
without $N$-body simulations). 

Research into the origin and longevity and/or transience of spiral
structure in disc galaxies has by and large relied on $N$-body
simulations
\citep{1971ApJ...168..343H,1977ARA&A..15..437T,sellwood89,sellwood91,
  2011MNRAS.410.1637S,ovh13,2014ApJ...785..137S} and on
(semi-)analytical mode analysis using the first-order CBE as a
starting point
\citep{1976PhDT........26Z,kalnajs77,polyfrid,palmer,b9,1997MNRAS.291..616P,1998MNRAS.300..106E,jalali05,jalali07,bintrem,poly15}. Here,
we use linear theory to investigate how gravitational softening affects
the growth of small-amplitude eigenmodes in $N$-body simulations.

We introduce the various softening techniques explored by us in
section \ref{sec:soft:2D}. Our implementation of gravitational
softening in linear mode theory is layed out in section
\ref{pystab}. The properties of the unsoftened axially symmetric disc
models are discussed in section \ref{themodel}. Our results are
presented in section \ref{sec:results} and their implications are
discussed in section \ref{sec:dis}.


\section{Gravitational softening in two spatial dimensions} \label{sec:soft:2D}
The usual motivation for softening in two-dimensional $N$-body
simulations is to account for the finite thickness of the stellar
disc, which is neglected in the razor-thin limit
\citep{1971Ap&SS..14...73M}. In this interpretation, the softening
length $\eps$ is no longer a numerical but a \emph{physical} parameter
and the modification of gravity and, consequently, of the dynamics is
deliberate, because real galaxies are not razor-thin
\citep{1992MNRAS.256..307R,1997A&A...324..523R}.

However, as we are interested in the errors introduced by the
softening-induced modification of gravity, we cannot adopt this
interpretation but must consider the razor-thin disc as the desired
physical model, whose modes one may attempt to recover with an
$N$-body simulation.

\subsection{Gravity bias} \label{sec:soft:error}

When inserting the softening kernel~\eqref{eq:soft:kern} into
equation~\eqref{eq:V:unsoft}, we obtain the softened potential
\begin{equation}
	\label{eq:V:soft}
	V(\vec{x}) \approx \hat{V}(\vec{x}) = \sum_i\frac{Gm_i}{\eps}
        \phi\left(\frac{|\vec{x}-\vec{x}_i|}{\eps}\right).
\end{equation}
Here, $\hat{V}(\vec{x})$ can be interpreted as an \emph{estimate},
based on the masses and positions of the simulation particles, for the
true potential $V(\vec{x})$. The mean-square error made by this
estimate can be decomposed into a variance,
\begin{equation}
  \mathrm{var}_{\vec{x}}(\hat{V}) = \left\langle \left[ \left\langle
  \hat{V}(\vec{x})\right\rangle-\hat{V}(\vec{x})\right]^2 \right\rangle,
\end{equation}
and a bias,
\begin{equation}
  \mathrm{bias}_{\vec{x}}(\hat{V}) = \left\langle\hat{V}(\vec{x})\right\rangle-V(\vec{x}).
  \end{equation}
Here, $\langle\cdot\rangle$ denotes the ensemble average of a quantity
over all $N$-body realisations, in the limit $N\to\infty$, of the
underlying smooth density distribution. The variance measures the mean
amplitude of the random fluctuations of a softened $N$-body potential
around its ensemble mean. In other words, it measures the graininess
of the $N$-body potential:~the error made by \emph{not softening
enough}. The bias measures the deviation of the ensemble mean of the
softened $N$-body potential from the underlying smooth potential. This
is the error made by \emph{softening too much}.

For the situation of three-dimensional $N$-body simulations,
\cite{2001MNRAS.324..273D} has derived analytical asymptotic relations
for these quantities. An adaptation of his derivations to an $N$-body
simulation of a razor-thin disc with surface density $\Sigma(\vec{x})$
gives
\begin{equation}
	\label{eq:bias:V:2D}
	\mathrm{bias}_{\vec{x}}(\hat{V}) =
        a_0\,\eps\,G\,\Sigma(\vec{x}) +a_2\,\eps^3
        G\,\vec{\nabla}^2\Sigma(\vec{x}) + \mathcal{O}(\eps^5),
\end{equation}
for the bias on the potential $V$ (see Appendix \ref{app1} for a
derivation). The coefficients $a_n$ depend only on the functional form of
the softening kernel:
\begin{equation}
  a_n =
  \frac{2\pi}{2^n\,([n/2]!)^2}\int_0^\infty\left[1-u\phi(u)\right]u^n\d
  u. \label{an}
\end{equation}

This is different for three-dimensional systems, where  this bias
asymptotes as $\eps^2$ at lowest order. Here, in two dimensions, the
gravity biases are proportional to $\eps$. This is a direct
consequence of the reduced number of dimensions. Thus, in
two-dimensional $N$-body simulations, the gravity bias is in general
significantly stronger than in three-dimensional $N$-body
simulations\footnote{\label{foot:bias:3D} Except in three-dimensional
  simulations of disc galaxies with scale height $h\apprle\eps$, since
  in three dimensions
\begin{subequations}
	\label{eq:bias:3D}
\begin{align}
	\mathrm{bias}_{\vec{x}}(\hat{V}) &\approx
        -a_0^{\mathrm{3D}}\eps^2G\rho(\vec{x})
        \\ \mathrm{bias}_{\vec{x}}(\hat{\vec{a}}) &\approx
        -a_0^{\mathrm{3D}}\eps^2G\vec{\nabla}\rho(\vec{x})\propto\eps^2/h
\end{align}
\end{subequations}
\citep[][eqs.~10]{2001MNRAS.324..273D} with $\rho(\vec{x})$ the
spatial density.}.

Note that the integrand in equation (\ref{an}) is always well-behaved
in the limit $u \to 0$ but can be problematic for large $u$. If
$1-u\phi(u) \propto u^{-p}$ for large $u$ then only coefficients $a_n$
with $n<p-1$ are finite; the rest come out infinite.  In this case,
the Taylor series~\eqref{eq:bias:V:2D} does not converge, but the
terms at $n<p-1$ still provide a useful approximation, only the
remainder grows faster than $\eps^{2\lceil (p-1)/2\rceil}$.

\begin{table*}{
	\caption{\label{tab:soft} Characteristics of the various
          softening methods used in this study. Here, $\phi$,
          $\varrho$, and $\sigma$ take argument $u=r/\eps$ and are the
          dimensionless kernels for, respectively, potential, spatial
          and surface density, defined in
          equations~\eqref{eq:soft:kern} and
          \eqref{eqs:soft:dens:surf}. The coefficients $a_0$ and $a_2$
          determine the gravitational biases (see
          equation~\ref{eq:bias:V:2D}). Here, $\varepsilon_0$
          quantifies the scaling of the kernels with non-zero $a_0$ to
          a common level of gravity bias and $\varepsilon_F$
          quantifies the scaling of the kernels to a common maximum
          inter-particle force. A dagger indicates softening methods
          with compact support, where the formul\ae{} for $\phi$ and
          $\varrho$ only apply at $r<\eps$ or
          $t\equiv1-u^2\in[0,1]$. For these methods, no sensible
          razor-thin surface-density kernel $\sigma(u)$ can be
          provided.  }
	\begin{tabular}{lllllcccc}
	\multicolumn{2}{c}{name}
	& $\phi(u)$ & $\varrho(u)$ & $\sigma(u)$ & $a_0$ & $a_2$ & $\varepsilon_0/\eps$ & $\varepsilon_F/\eps$ \\
	\hline
	& Newton	& $\displaystyle\frac{1}{u}$
			& $\displaystyle\delta_{\mathrm{3D}}(u)$
			& $\displaystyle\delta_{\mathrm{2D}}(u)$
			& 0 & 0 & 0 & 0
		\\[2ex]
	P$_0$ & Plummer & $\displaystyle\frac{1}{\sqrt{1+u^2}}$
			& $\displaystyle\frac{3}{4\pi}\frac{1}{(1+u^2)^{5/2}}$
			& $\displaystyle\frac{1}{2\pi}\frac{1}{(1+u^2)^{3/2}}$
			& $2\pi$ & $\infty$ & 1 & 1
		\\[3ex]
	Q$_2$ & 
	$\begin{array}{l}\text{2$^\mathrm{nd}$\,modified}\\ \text{Kuz'min}\end{array}$
			& $\displaystyle\frac{3+\tfrac52u^2+u^4}{(1+u^2)^{5/2}}$
			& $\displaystyle\frac{15}{8\pi}\frac{4-3u^2}{(1+u^2)^{9/2}}$
			& $\displaystyle\frac{3}{4\pi}\frac{4-u^2}{(1+u^2)^{7/2}}$
			& 0 & $\displaystyle-\frac{\pi}{3}$ & --- & 2.568
		\\[3ex]
	$^{\dag}$F$_3$ & Ferrers $n=3$
			& $1+\frac12t+\frac38t^2+\frac5{16}t^3+\frac{35}{128}t^4$
			& $\displaystyle \frac{315}{64\pi}t^3$
			& ill-behaved
			& $\displaystyle\frac{63\pi}{128}$ 
			& $\displaystyle\frac{7\pi}{1024}$ 
			& $\displaystyle\frac{63}{256}$ & 2.309
			\\[3ex]
	$^{\dag}$L$_2$ & 
		$\begin{array}{l}\text{2D modified}\\ \text{Ferrers $n=2$}\end{array}$
			& $1+\frac12t+\frac38t^2+\frac{5}{2}t^3$
			& $\displaystyle\frac{105}{8\pi}(1-2u^2)t$
			& ill-behaved
			& $\displaystyle0$ 
			& $\displaystyle-\frac{7\pi}{384}$ 
			& --- & 3.711
			\\
	\hline
	\end{tabular}
  }
\end{table*}

\subsection{Softening kernels} \label{sec:soft:kernels}
\begin{figure}
\includegraphics[trim=15 15 10 10,clip,width=0.475\textwidth]{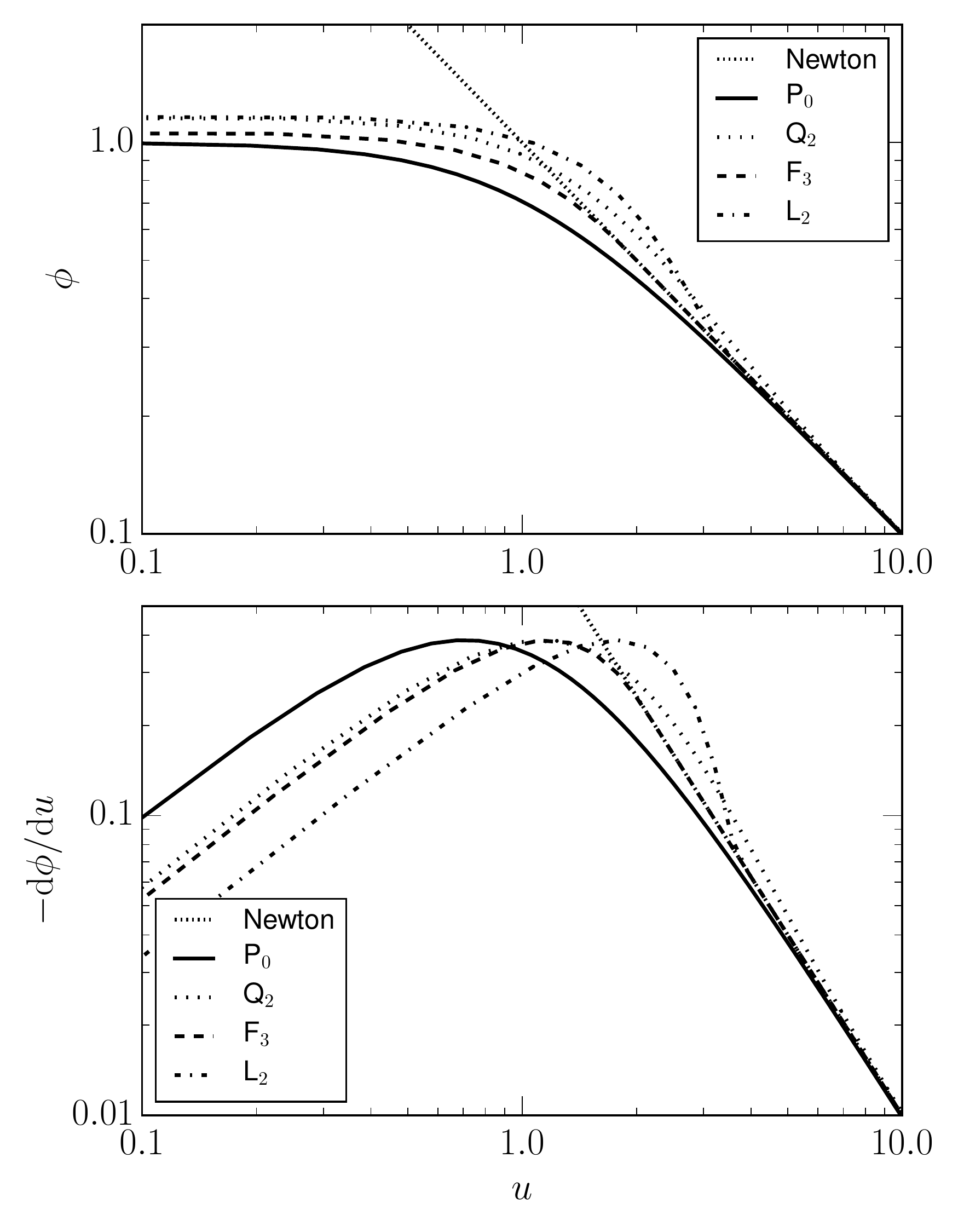}
\caption{Top panel:~the dimensionless interaction potential of the
  softening recipes listed in Table~\ref{tab:soft}. Bottom panel:~the
  corresponding interparticle forces. The dimensionless interaction
  potentials (top) and inter-particle forces (bottom) of these
  softening kernels are listed in Table \ref{tab:soft}. The kernels
  are scaled according to equations~\eqref{eq:eps:scaling} and
  \eqref{eq:eps:scale:force} to obtain a common maximum inter-particle
  force (as is obvious in the bottom panel), see also Section
  \ref{sec:soft:scale}. Note that kernels with $a_0=0$ (Q$_2$ and
  L$_2$) have super-Newtonian gravity at intermediate scales to
  compensate for the sub-Newtonian behaviour at $r\to0$.
 \label{fig:Id}}
\end{figure}
The functional forms and other properties of the softening kernels
used in this study are listed in Table~\ref{tab:soft}. Their
interparticle interaction potentials and forces are plotted in Figure
\ref{fig:Id}. In particular, we list the dimensionless interparticle
interaction potential $\phi$ and the corresponding three-dimensional
and two-dimensional dimensionless density kernels, denoted by
$\varrho$ and $\sigma$, respectively, assigned to each point particle.

In three dimensions, gravitational softening is equivalent to
estimating the spatial mass density as a superposition of spheres with
density distribution $\rho(r)$ placed at the particle positions. In
two dimensions, one can consider softening as a way of smoothing the
overall mass distribution as a superposition of razor-thin discs with
density distribution $\Sigma(r)$ at the particle positions. For a
softening kernel $\phi$, these spherical or razor-thin surface density
distributions are given by
\begin{equation}
	\rho(r) = \frac{m}{\eps^3} \varrho\left(\frac{r}{\eps}\right)
        \qquad\text{and}\qquad \Sigma(r) = \frac{m}{\eps^2}
        \sigma\left(\frac{r}{\eps}\right), \label{eq:soft:rS}
\end{equation}
where the dimensionless density and surface-density kernels are given
by
\begin{subequations}
	\label{eqs:soft:dens:surf}
\begin{align}
	\label{eq:soft:dens:kernel}
	\varrho(u) &= -\frac{1}{4\pi u^2}\frac{\d}{\d
          u}\left(u^2\frac{\d\phi}{\d u} \right), \\[1ex]
	\label{eq:soft:surf:kernel}
	\sigma(u) &= -\frac{1}{\pi^2} \int_{u}^\infty\frac{\d
          x\,}{\sqrt{x^2-u^2}} \frac{\d}{\d x}\int_0^{x} \frac{x\,\d
          t}{\sqrt{x^2-t^2}}\,\frac{\d\phi(t)}{\d t},
\end{align}
\end{subequations}
respectively \citep{bintrem,1999ASPC..165..325K}. 

Note that the softened force is that between a softened particle (with
density distribution $\rho(r)$ or $\Sigma(r)$ given by
eqn. (\ref{eq:soft:rS})) and a point particle and {\em not} the force
between two softened particles \citep{2012MNRAS.425.1104B}. This
remark applies to all softening techniques.

\subsubsection{Plummer softening P$_0$:~infinite support, $a_0 \ne 0$} \label{sec:soft:P0}

In three dimensions, this popular softening method corresponds to
estimating the spatial mass density as a superposition of Plummer
spheres \citep{1911MNRAS..71..460P} with scale radius $\eps$ at the
particle positions. In two dimensions, Plummer softening amounts to
smoothing the overall mass distribution as a superposition of
razor-thin Kuz'min discs \citep{Kuzmin56}.

Plummer softening modifies the gravitational interaction at all
interparticle separations and asymptotes to the Newtonian interaction
only for infinitely large particle separations. As a result,
$a_0^{\mathrm{3D}}=\infty$ and the gravity bias~\eqref{eq:bias:3D} in
this case grows faster than $\propto\eps^2$. In two dimensions, this
method's non-zero $a_0$ indicates that the gravity bias increases
linearly with softening length.

\subsubsection{Modified Kuz'min softening Q$_2$:~infinite support, $a_0 =0$} \label{sec:soft:Q2}

One can modify the Plummer kernel such that $a_0$, $a_2$, {\ldots},
$a_k=0$, for any chosen even $k$, in order to significantly reduce the
gravity bias. For instance, we introduce the class of modified Kuz'min
potentials, with an interaction potential given by
\begin{equation}
  \phi(u) = \frac{K_n(u^2)}{\left( 1+u^2 \right)^{n+1/2}},
\end{equation}
here $K_n(u^2) = \sum_{i=0}^n c_i u^{2i}$ is a polynomial of degree
$n$ in $u^2$ with coefficients $c_i$. We refer to the $n^{\rm th}$
member of this class as Q$_n$. Clearly, the choice $n=0$ yields the
Plummer kernel, or:~Q$_0$=P$_0$.

The extra degree of freedom that comes with the choice $n=1$, can be
exploited to make $a_0=0$. For instance, for this Q$_1$ kernel, one
finds that
\begin{align}
  a_0 &\propto \lim_{x \rightarrow \infty} \int_0^x \left( 1 -
  \frac{c_0 u + c_1 u^3}{\left( 1+u^2 \right)^{3/2}} \right)\mathrm{d}u \nonumber \\
  & =\lim_{x \rightarrow \infty} \left\{
  (1-c_1)x + 2c_1-c_0 + {\cal O}\left( \frac{1}{x} \right) \right\}
\end{align}
and the choice $c_0=2$, $c_1=1$ makes $a_0=0$. Thus, the `Q$_1$'
method is defined by the interparticle potential \beqn \phi(u) =
\frac{2+u^2}{(1+u^2)^{3/2}} \neqn and by the corresponding 3D and 2D
density distributions
\begin{equation}
	\varrho(u) = \frac{3}{4\pi}\frac{4-u^2}{(1+u^2)^{7/2}}
        \,\,\text{ and }\,\, \sigma(u) =
        \frac{3}{2\pi}\frac{1}{(1+u^2)^{5/2}}. \label{eq:soft:Q1}
\end{equation}
In order to achieve $a_0=0$, this softening
method compensates with slightly super-Newtonian forces at $r\apprge
1.2\eps$ for the substantially sub-Newtonian accelerations at small
separations. Unfortunately, while we now have $a_0=0$, the Q$_1$
softening kernel still has a diverging second-order coefficient $a_2$,
such that the gravity bias grows faster than $\eps^3$.

The extra degree of freedom provided by coefficient $c_2$ of kernel
Q$_2$ can be used to provide $a_2$ with a finite value. Indeed, for
this kernel, we find that
\begin{align}
  a_0 &\propto \lim_{x \rightarrow \infty} \int_0^x \left( 1 -
  \frac{c_0 u + c_1 u^3+ c_2 u^5}{\left( 1+u^2 \right)^{5/2}} \right)\mathrm{d}u \nonumber \\
  &= \lim_{x \rightarrow \infty} \left\{ (1-c_2)x + \frac{1}{3} ( 8c_2-2c_1-c_0) + {\cal O}\left( \frac{1}{x} \right)
  \right\}
\end{align}
and
\begin{align}
  a_2 &\propto \lim_{x \rightarrow \infty} \int_0^x \left( 1 -
  \frac{c_0 u + c_1 u^3+ c_2 u^5}{\left( 1+u^2 \right)^{5/2}}  \right)u^2\mathrm{d}u \nonumber \\
  &= \lim_{x \rightarrow \infty} \left\{ \frac{1}{3}(1-c_2)x^3 + \left( \frac{5c_2}{2}-c_1\right)x
  \right. \nonumber \\ & \left. \hspace{8em}- \frac{2}{3} ( 8c_2-4c_1+c_0 ) 
  + {\cal O}\left( \frac{1}{x} \right)\right\}.
\end{align}
Demanding $a_0$ to be zero and $a_2$ to be finite, leads to $c_2=1$,
$c_1=5/2$, and $c_0=3$. All properties of this Q$_2$ kernel are listed
in Table~\ref{tab:soft}.

Of course, this game can be continued to obtain $a_2=0$, then finite
$a_4$, then $a_4=0$, etc. However, the functional form of the
resulting interaction potentials becomes increasingly complex. 

\subsubsection{Ferrers $n=3$ softening F$_3$:~compact support, $a_0 \ne 0$} \label{sec:soft:F3}\label{sec:soft:cubic:spline}

An increasingly popular method is cubic spline softening, which in
three-dimensional $N$-body simulations corresponds to replacing each
particle by a cubic-spline smoothing kernel as is widely used in
Smoothed Particle Hydrodynamics (SPH) codes \citep[see
  e.g.][]{1992ARA&A..30..543M} and was introduced as a gravitational
softening kernel for $N$-body/SPH codes by
\cite{1989ApJS...70..419H}. Its main advantages in this context are
its exactly Newtonian behaviour beyond the softening length and its
dual use as hydrodynamics smoother and gravity softener. However, its
interparticle potential and 3D density distribution are numerically
rather unattractive due to their complex, piecewise continuous
functional forms.

Here and in the remainder, we will use the phrase ``compact support''
to indicate that a kernel yields exactly Newtonian forces outside the
softening kernel. In three dimensions, it immediately follows from
Newton's first theorem that the corresponding density distribution is
zero outside the kernel, i.e. $\varrho=0$ at $u>1$. In two dimensions,
this is not the case. In fact, all softening kernels with exact
Newtonian gravity at separations $r>\eps$ have poorly behaved
corresponding razor-thin disc profiles
$\sigma(u)$~\eqref{eq:soft:surf:kernel}, with infinite spatial extent
and negative values.

We here opt for the so-called Ferrers softening methods, labeled
`F$_n$', whose interaction potentials are polynomials of degree $n+1$
in the variable $t=1-u^2$ inside the softening length and that behave
exactly Newtonian at separations $r>\eps$. In three dimensions, they
correspond to replacing each particle with a \cite{Ferrers} sphere of
order $n$. For $n=0$ this is just a homogeneous sphere. Higher-order
models have spherical densities that are simple polynomials, with $n$
continuous derivatives, in the variable $t=1-u^2$.

For this paper, we investigate member $n=3$ from this family, with
properties listed in Table~\ref{tab:soft}, as an example of a
softening method with compact support but with $a_0\ne 0$.

\subsubsection{Modified Ferrers softening L$_2$:~compact support, $a_0 = 0$} \label{sec:soft:L1}

It is possible to modify the Ferrers softening methods to obtain
$a_0=0$ in order to reduce their gravitational bias while retaining
the attractive property of having compact support. Here, we test the
method labeled 'L$_2$', or `2D modified Ferrers $n=2$', in
Table~\ref{tab:soft}. Its name derives from the fact that it's based
on the F$_2$ kernel, which has as an interaction potential
$\phi(u)=1+\frac{1}{2}t+\frac{3}{8}t^2+\frac{5}{16}t^3$, but where the
coefficient of the last term is tuned to make $a_0=0$. This leads to
the L$_2$ interaction potential
$\phi(u)=1+\frac{1}{2}t+\frac{3}{8}t^2+\frac{5}{2}t^3$.

This kernel achieves its desirable properties by
having super-Newtonian accelerations for a limited range of
separations close to and inside of $r=\eps$.

\subsection{Softening scale} \label{sec:soft:scale}

The softening length and kernel are only defined up to a
re-scaling:~the softened potential \eqref{eq:V:soft} is invariant under the
transformation
\begin{equation}
\label{eq:eps:scaling}
\varepsilon \to a\varepsilon
\qquad\text{and}\qquad
\phi(q) \to a\phi(aq)
\end{equation}
with \emph{scaling factor} $a$. This implies that the parameter
$\varepsilon$ has no natural scale by itself and comparing different
kernels at the same $\varepsilon$ is meaningless. Therefore, some other
measure is required for such a comparison. One such measure valid for
all softening kernels is the force-scaling as
\begin{equation}
\label{eq:eps:scale:force}
\varepsilon_{F} = \varepsilon / \sqrt{-\phi'_{\max}},
\end{equation}
where $-\phi'_{\max}$ denotes the maximum value of the derivative of
$\phi$ \citep{2010MNRAS.401..791S}. The ratios $\varepsilon_{F}/\varepsilon$, scaled to unity for the
Plummer kernel, are listed in Table \ref{tab:soft} for the kernels
considered in this study.

For kernels with $a_0\neq0$, another natural measure of the softening
length is
\begin{equation}
	\label{eq:soft:scale}
	\varepsilon_0\equiv \frac{a_0}{2\pi}\eps =
        \int_0^\infty\left[1-r\pot(r)\right]\d r,
\end{equation}
which measures the actual scale of the bias irrespective of any
re-scaling. With this definition, $\varepsilon_0=\eps$ for Plummer
softening. For other softening methods used in this study,
$\varepsilon_0$ is given in Table~\ref{tab:soft}.

Likewise, softening techniques with zero $a_0$ but non-zero $a_2$ can
be scaled to a common level of gravity bias via the softening length
transformation
\begin{equation}
  \varepsilon_2 = \left|
  \frac{3a_2}{\pi} \right|^{1/3} \varepsilon. \label{a2scal}
\end{equation}
With this definition, $\eps=\varepsilon_2$ for Q$_2$ softening.

\section{Softened gravity in stability analysis}
\label{pystab}

\subsection{Linear mode theory} \label{linmodthe}

We use {\sc pyStab}, a {\sc Python}/{\sc C++} computer code, to
analyse the stability of a razor-thin stellar disc embedded in an
axially symmetric gravitational potential. The details of the
mathematical formalism behind this code and of its implementation can
be found in \citet{b9}, \citet{dury08}, and \citet{dv16} so we will
not repeat these here. An axially symmetric disc galaxy model is
characterized by a distribution function $F_0(E,J)$, with $E$ the
specific binding energy and $J$ the specific angular momentum of a
stellar orbit, and a mean gravitational potential $V_0(r)$. In the
remainder, we will refer to this unperturbed axially symmetric state
as the ``base state'' of the system. Note that this base state is only
a correct solution of the CBE when employing the Newtonian
gravitational interaction (but see below).

For any given base state, {\sc pyStab} can retrieve those complex
frequencies $\omega$ for which a spiral-shaped perturbation of the
form
\begin{equation}
	V_{\rm pert}(r,\theta,t) = V_{\rm
          pert}(r)\,\mathrm{e}^{\mathrm{i}( m\theta-\omega t)}
\end{equation}
constitutes an eigenmode. Here, $(r,\theta)$ are polar coordinates in
the stellar disc, $m$ is the multiplicity of the spiral pattern,
$\Omega_p=\Re\{\omega\}/m$ its pattern speed, and $\Im\{\omega\}$ its
growth rate. A general perturbing potential can always be expanded in
such modes and, owing to the linear approximation, these can be
studied independently from each other.

In essence, {\sc pyStab} solves the first-order CBE to find the
response distribution function $f_{\rm resp}(r,\theta, v_r,
v_\theta,t)$ produced by a given perturbation $V_{\rm
  pert}(r,\theta,t)$. This response distribution function generates
the response density
\begin{equation}
	\Sigma_{\rm resp}(r,\theta,t) = \int \sub{f}{resp}(r,\theta,
        v_r, v_\theta,t) \,\d v_r\d v_\theta
\end{equation}
which in turn {gives rise to the} response {softened} gravitational
potential
\begin{equation}
	\label{Vprp}
	V_{\rm resp}(\vec{x}) = G \int\Sigma_{\rm
          resp}(\vec{x}')\,\pot(|\vec{x}-\vec{x}'|)\,\d^2\!\vec{x}',
\end{equation}
where the integral runs over the whole surface of the stellar disc
{and $\pot(r)$, defined in equation~\eqref{eq:soft:kern}, is the
  softened Green's function for} gravitational interactions{,
  replacing the Newtonian $1/r$}.
Eigenmodes are then identified by the fact that
\begin{equation}
	V_{\rm pert}(r,\theta,t) \equiv V_{\rm resp}(r,\theta,t)
\end{equation}
and \textsc{pyStab} employs a matrix method \citep{kalnajs77} to find
them. The perturbing potential $V_{\rm pert}$ is expanded in a basis
of potentials, $V_\ell$. The response to each basis potential, denoted
by $V_{\ell,\rm resp}$, can likewise be expanded in this basis as
\begin{equation}
	V_{\ell,\rm resp} = \sum_{k} {\mathcal C}_{k \ell} V_{k}.
\end{equation}
If the perturbation is an eigenmode, then the ${\mathcal C}$ matrix
can be shown to possess a unity eigenvalue
\citep{b9,dury08,dv16}. This feature is exploited by {\sc pyStab} to
identify the eigenmodes.

The formalism contains a number of technical parameters, such as the
number of orbits on which phase space is sampled (here we use $n_{\rm
  orbit}(n_{\rm orbit}+1)/2$ orbits with $n_{\rm orbit}=600$ in the
allowed triangle of turning point -- or pericentre/apocentre --
space), the number $n_{\rm Fourier}$ of Fourier components in which
the periodic part of the perturbing potential is expanded (here we use
$n_{\rm Fourier}=80$), the number of potential-density pairs (PDPs)
that is used for the expansion of the radial part of the perturbing
potential and density (we use 44 PDPs), and the shape and extent of
the PDP density basis functions. As in \citet{dv16}, we use PDP
densities of the form
\begin{equation}
	\label{basefunc}
	\Sigma_\ell(r) = \Sigma_0(r)
        \exp\left(-\frac{1}{2}\left(\frac{r-r_\ell}{\sigma_\ell}\right)^2\right)
\end{equation}
where the average radii $r_\ell$ cover the relevant part of the
stellar disc and are evenly spaced on a logarithmic scale so the
resolution is highest in the inner regions of the disc. The widths
$\sigma_\ell$ are automatically chosen such that consecutive basis
functions are sufficiently unresolved to represent any smooth
function. The position of these PDP density basis functions can be
tuned to achieve a high spatial resolution there where the eigenmodes
live. The corresponding PDP potentials are obtained via
\begin{equation}
	\label{Vpdp}
	V_{\ell}(\vec{x}) = G \int \Sigma_{\ell}(\vec{x}')
        \,\pot(|\vec{x}-\vec{x}'|)\,\d^2\!\vec{x}'.
\end{equation}

\subsection{Introducing gravitational softening}

Since we want to validate our approach by comparing particular results
with published work based on numerical simulations, we mimic the
strategies employed by simulators when setting up and performing
$N$-body simulations of disc galaxies aimed at mode analysis. Usually,
an initial condition is generated by sampling stellar particles from
the distribution function $F_0(E,J)$ evaluated using the Newtonian
gravitational potential $V_0(r)$, independent of the gravitational
softening that is employed later on when evolving the particles
through time. Moreover, the axially symmetric force field of the base
state is subsequently evaluated correctly, i.e. without softening,
either by directly using the analytical expression for the potential
$V_0$ or by adding a small correction to the softened gravitational
field derived from the particles. Only the non-axisymmetric force
field of the growing waves is softened
\citep{1995ApJ...451..533E,2001ApJ...546..176S,2012ApJ...751...44S}.
This allows a simulator to sample particles from the correct DF
evaluated in the correct potential so that at least the initial
conditions of a simulation correspond to the intended base state and
the particle dynamics in the axially symmetric force field is followed
correctly.

Therefore, we only implement gravitational softening in the response
potential $V_{\rm resp}(r,\theta,t)$, but not in the axially symmetric
base state potential $V_0(r)$. Using this strategy, eqn. (\ref{Vprp}),
and a fortiori eqn (\ref{Vpdp}), is the only place where the softened
gravitational interaction enters the computation of the modes. It is
therefore straightforward to insert interaction potentials other than
the Newtonian one into a mode analysis code. The gravity bias
introduced in Section \ref{sec:soft:error} must then be regarded as a
measure for the fidelity with which the softened response potential
resembles the Newtonian one. Thus, we can use linear stability theory
to emulate the results expected in the large $N$ limit from $N$-body
simulations of disc galaxies. In this paper, we investigate the effect
on the eigenmodes in disc galaxy models from the P$_0$, Q$_2$, F$_3$,
and L$_2$ softening methods listed in Table~\ref{tab:soft}.

\section{The base states}\label{themodel}

Below, we give the essential details of the two base states that we
employ for this study. We also list the frequencies of the known
eigenmodes of these base states computed for a Newtonian interparticle
interaction.

\subsection{The isochrone disc model} \label{isochrone}

The isochrone disc is characterized by the cored density profile
\begin{equation}
	\Sigma_0(r) = \frac{Mb}{2\pi r^3}\left( \ln
        \frac{r+\sqrt{r^2+b^2}}{b} -\frac{r}{b} \right)
\end{equation}
which self-consistently generates the gravitational potential
\begin{equation}
	V_0(r) = \frac{GM}{b+\sqrt{b^2+ r^2}},
\end{equation}
\citep{1959AnAp...22..126H,1976ApJ...205..751K}. Here, $M$ is the
total mass of the stellar disc and $b$ its scale-length. As shown by
\cite{1976ApJ...205..751K} {and} \citet{1995ApJ...451..533E}, a family
of distribution functions that generate this potential-density pair
is given by
\begin{equation}
	F_0(E,J) = \left[ \frac{E}{V_0(0)} \right]^{{m_K}-1}g_{m_K}(x)
\end{equation}
with $m_K$ an integer, $x = J\sqrt{2E}/GM$, and
\begin{align}
	g_{m_K}(x) &= \frac{2^{m_K}}{2\pi V_0(0)}\Bigg[x\frac{\d
            \tau_{m_K}}{\d x} - \frac{{m_K}({m_K}-3)}{2} \tau_{m_K}(x)
          \nonumber \\ &\phantom{=\frac{2^{m_K}}{2\pi V_0(0)}\Bigg[} +
            \int_0^1 \tau_{m_K}(\eta x)\,\eta^{m_K}\frac{\d
              ^2P_{{m_K}-1}}{\d \eta^2} \d\eta \Bigg].
\end{align}
Here, $P_m$ is the Legendre polynomial of degree $m$ and
\begin{align}
	\tau_{m_K}(x) &= -\frac{M}{16\pi
          b^2}\frac{(1-x^2)^{3-{m_K}}}{x^3(1+x^2)} \nonumber
        \\ & \hspace{4em} \times \left[ 2x + (1+x^2) \ln\frac{1-x}{1+x}
          \right].
\end{align}
We adopt $m_K=12$ for this study. The Legendre polynomial can be
evaluated explicitly, allowing the integral featuring in the
expression for the distribution function to be evaluated in closed
form. However, the resulting expression is numerically very unstable
for small $x$. Therefore, we opted to simply evaluate the integral
numerically.  This distribution function is only used to populate
orbits with positive angular momentum.

Counter-rotating stars have been added according to the prescription
given in \citet{1995ApJ...451..533E} which, unfortunately,
necessitates going back and forth between energy, angular momentum and
the radial action variable, $J_r$:
\begin{equation}
	F'_0(E,J) =\left\{ \begin{array}{ll} \frac12 F_0(E',0) \qquad
          & \text{if $J<0$} \\[1ex] F_0(E,J) - \frac12 F_0(E',0)
          \qquad & \text{if $J>0$}
\end{array}
\right.
\end{equation}
with $E'$ the energy corresponding to a radial action $J_r+|J|$ and
zero angular momentum. Fortunately, analytical conversion formul\ae{}
between energy, angular momentum, and radial action exist for the
isochrone disc \citep{bintrem}.

We will focus here on the bisymmetric ($m=2$) modes of this
model. \citet{1997MNRAS.291..616P} provide the frequencies of three
modes of this base state model, choosing units such that $G=M=b=1$, as
\begin{equation}
\begin{array}{r@{=}r@{\,+\,}r@{\,\mathrm{i}}}
	\omega_1 & 0.59 & 0.21 \\[0.5ex] \omega_2 & 0.46 & 0.14
        \\[0.5ex] \omega_3 & 0.26 & 0.05
\end{array}
\end{equation}
while \citet{jalali05} find
\begin{equation}
\begin{array}{r@{=}r@{\,+\,}r@{\,\mathrm{i}}}
  \omega_1 & 0.584 & 0.217 \\[0.5ex] \omega_2 & 0.468 & 0.148
\end{array}
\end{equation}
for the two main modes\footnote{\label{foot:acc} Accuracy estimates
  for mode frequencies derived from linear stability computations are
  hard to obtain since many numerical parameters come into
  play. Judging from the differences between published mode
  frequencies and from our own limited experiments with varying the
  values of the employed numerical parameters (described in Section
  \ref{linmodthe}) we estimate the mode frequencies to be accurate to
  about the percent level.  }.

\subsection{The Mestel disc} \label{mestel}

The \citet{1963MNRAS.126..553M} disc has a cusped total surface
density given by
\begin{equation}
	\Sigma_0(r) = \Sigma_0 \frac{r_0}{r}
\end{equation}
which self-consistently generates a gravitational potential of the
form
\begin{equation}
	V_0(r) = -v_0^2\ln \left( \frac{r}{r_0} \right)
\end{equation}
with the surface density scale given by $\Sigma_0 = v_0^2/2\pi G r_0$.
Here, $v_0$ is the value of the disc's constant circular velocity. A
central hole is cut out of this disc model by multiplying its
distribution function \citep{1977ARA&A..15..437T}
\begin{equation}
	f(E,J) = \frac{\Sigma_0 v_0^q}{\sqrt{2^q \pi} \Gamma\left(
          \frac{q+1}{2}\right) \sigma^{q+2} } \left( \frac{J}{r_0 v_0}
        \right)^q \mathrm{e}^{E/\sigma^2},
\end{equation}
where $q$ is a real number, with a cut-out function of the form
\begin{equation}
	H_{\rm cut}(J) = \frac{x}{1+x}
\end{equation}
with $x=\left( J/r_0 v_0 \right)^n$ (obviously, this also slightly
suppresses the distribution function at larger $J$-values). Outside
the central cut-out region, the disc's constant radial velocity
dispersion is given by $\sigma = v_0/\sqrt{1+q}$.


Here, we will focus on the $q=6$, $n=4$ member of this model family
and adopt units such that $G=v_0=r_0=1$. Its dominant bisymmetric mode
is then expected to have a frequency
\begin{equation}
	\omega_1 \approx 0.88+0.13\,\mathrm{i},
\end{equation}
as shown, e.g., by
\citet{1977ARA&A..15..437T,1997PhDT.........1R,1998MNRAS.300..106E,poly15}. This
mode owes its existence to the inner cutout:~if the angular momentum
cut-off is not sufficiently steep, i.e. if $n$ is too small, there is
no eigenmode. The idea is that incoming trailing wave packets are
(partially) reflected from this sharp inner edge to travel back
outwards as leading waves. Overreflection, or swing amplification, at
the evanescent zone around the corotation resonance
\citep{1976ApJ...205..363M,toomre81} sends amplified trailing waves
back inwards. Inside this resonance cavity, growing modes can occur
\citep{1998MNRAS.300..106E}. 
\section{Results} \label{sec:results}


\subsection{The isochrone disc}
\subsubsection{Unsoftened gravity} \label{bigL}

For the two most rapidly growing $m=2$ modes of the $m_K=12$ isochrone
disc, we find frequencies
\begin{equation}
\begin{aligned}
  \omega_1 &= 0.582+0.215\;\mathrm{i}\\
  \omega_2 &= 0.466+0.146\;\mathrm{i},
\end{aligned}
\end{equation}
in good agreement with published values.

However, the third mode listed by \citet{1997MNRAS.291..616P} showed
up in our analysis as only the fifth fastest growing mode, with a
frequency
\begin{equation}
\begin{array}{r@{=}r@{\,+\,}r@{\;\mathrm{i}}}
	\omega_3 & 0.272 & 0.053
\end{array}.
\end{equation}
The two interloping modes at
\begin{equation}
\begin{array}{r@{=}r@{\,+\,}r@{\;\mathrm{i}}}
	\omega'_3 & 0.384 & 0.103 \\[0.5ex] \omega'_4 & 0.323 & 0.075
\end{array}	
\end{equation}
have not been described in the literature before. We confirmed that
they are robust to changes of the numerical parameters in the code
(resolution in phase space, number of Fourier modes, etc.) and that
they exert a zero total torque on the disc, as they should, and
therefore see no reason to discard them as spurious \citep{poly15}.

\subsubsection{Softened gravity}

\begin{figure}
	\includegraphics[trim=0 12.5 0
          0,clip,width=0.49\textwidth]{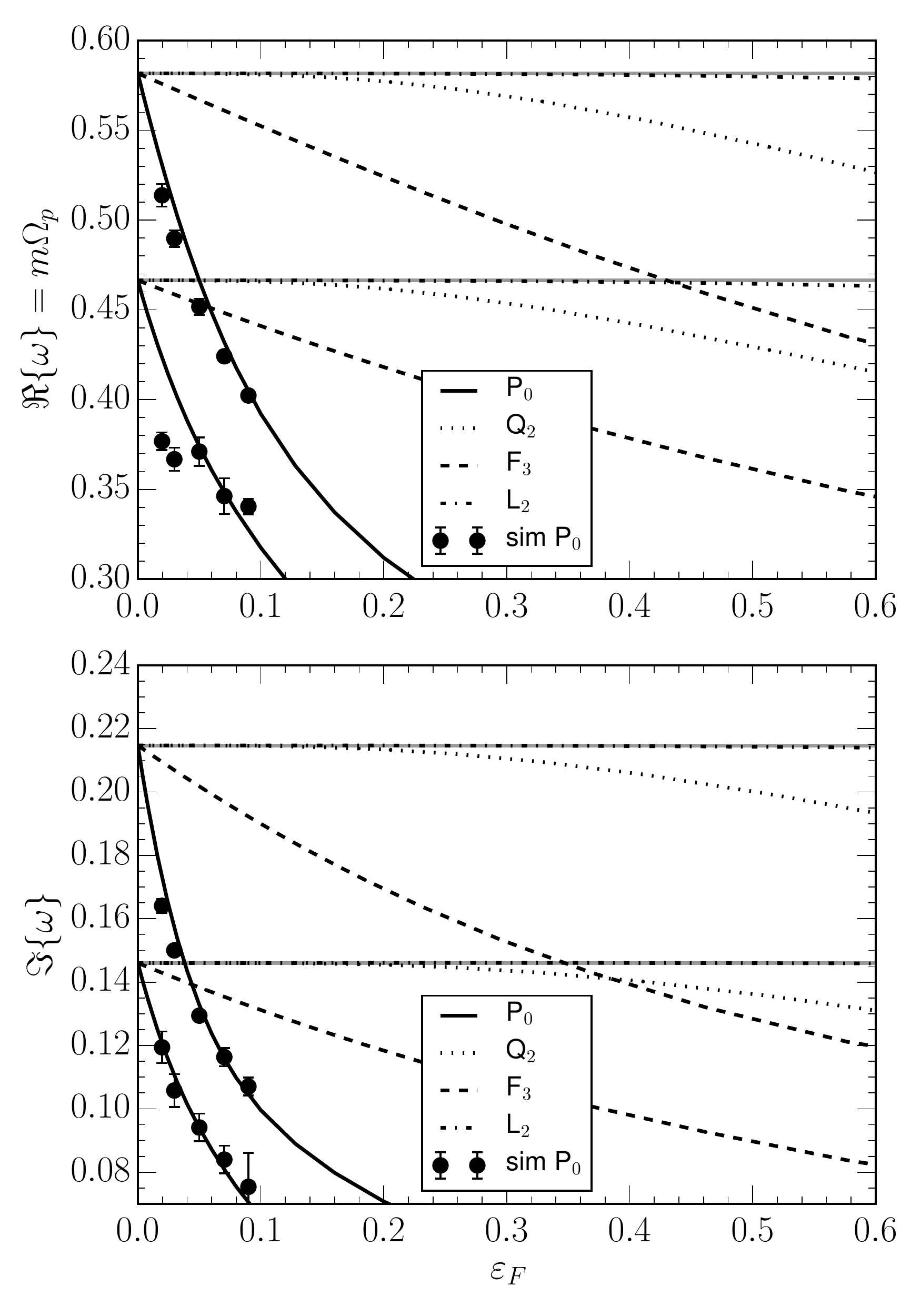}
	\caption{Pattern speeds $\Re\{\omega\}=m\Omega_p$ (top panel)
          and growth rates $\Im\{\omega\}$ (bottom panel) of the two
          dominant $m=2$ modes of the $m_K=12$ isochrone disc, with
          unsoftened frequencies of $\omega_1=0.582+0.215\,{\mathi}$
          and $\omega_2=0.466+0.146\,{\mathi}$ for different softening
          recipes (Plummer P$_0$, modified Kuz'min Q$_2$, Ferrers
          F$_3$, and modified Ferrers L$_2$). The softening lengths
          are scaled according to equations~\eqref{eq:eps:scaling} and
          \eqref{eq:eps:scale:force} to obtain a common maximum
          inter-particle force. The bullets are data taken from the
          $N$-body simulations using Plummer softening reported in
          \citet{1995ApJ...451..533E}. The horizontal grey lines indicate the
          Newtonian mode frequency.
          \label{fig:Isochrone_allsoft}}
\end{figure}

For this base state, published information on how the properties of
the two main $m=2$ modes change with softening in a numerical
simulation exists. \citet{1995ApJ...451..533E} use a polar grid code
with 120,000 particles, a fixed time step, and a polar grid of 128
azimuthal and 85 radial nodes to simulate the $m_K=12$ isochrone disc
from quiet-start initial conditions \citep{1983JCoPh..50..337S} using
Plummer softening with different softening lengths. The results of
these simulations are shown in Fig. \ref{fig:Isochrone_allsoft} as
black data points. Both the pattern speed and the growth rate of the
two dominant modes appear to be declining functions of softening
length. As \citet{1995ApJ...451..533E} note:~``[\ldots] it is clear
that both parts of the eigenfrequency are strongly affected by even
moderate softening.'' The dependence of the mode frequencies on
softening length is markedly non-linear which hampers a simple
extrapolation to zero softening length.

Overplotted in Fig. \ref{fig:Isochrone_allsoft} are the results from
our linear stability analysis with {\sc pyStab}, using different
softening prescriptions. Clearly, our results for Plummer softening
agree rather well with those presented in
\citet{1995ApJ...451..533E}:~both the pattern speed and growth rate
are non-linearly declining functions of the softening length (scaled
according to equations~\eqref{eq:eps:scaling} and
\eqref{eq:eps:scale:force} to a common maximum inter-particle
force). The drop is steepest for small softening lengths and becomes
shallower for larger $\eps$. Especially for larger $\eps$-values,
numerical simulations and linear mode analysis predict the same
behaviour for frequency as a function of softening length. We
tentatively attribute the deviations between the simulations and
linear theory at small $\eps$-values to variance, i.e.~to the gravity
error caused by not softening enough (cf. paragraph
\ref{sec:soft:error}).

Using the other softening recipes, the pattern speed and growth rate
likewise decline with increasing softening length but they do so much
less dramatically and with smaller deviation from a linear dependence
on softening length than when using Plummer softening. Moreover, it
appears that having a zero gravity bias parameter $a_0$ induces a much
stronger effect than having compact support. This is exemplified in
this case by the Q$_2$ method (infinite support, $a_0=0$) yielding
results much closer to the Newtonian ones than the F$_3$ method
(compact support, $a_0>0$). Methods that combine compact support with
having $a_0=0$, like the L$_2$ method, appear vastly superior, with
very little deviation between the retrieved mode frequencies and the
correct, Newtonian values.

However, we refer the reader to section \ref{sec:dis} for a discussion
of how to correctly interpret this apparent success.



\subsection{The Mestel disc}


\begin{figure}
\includegraphics[trim=0 12.5 0
  0,clip,width=0.49\textwidth]{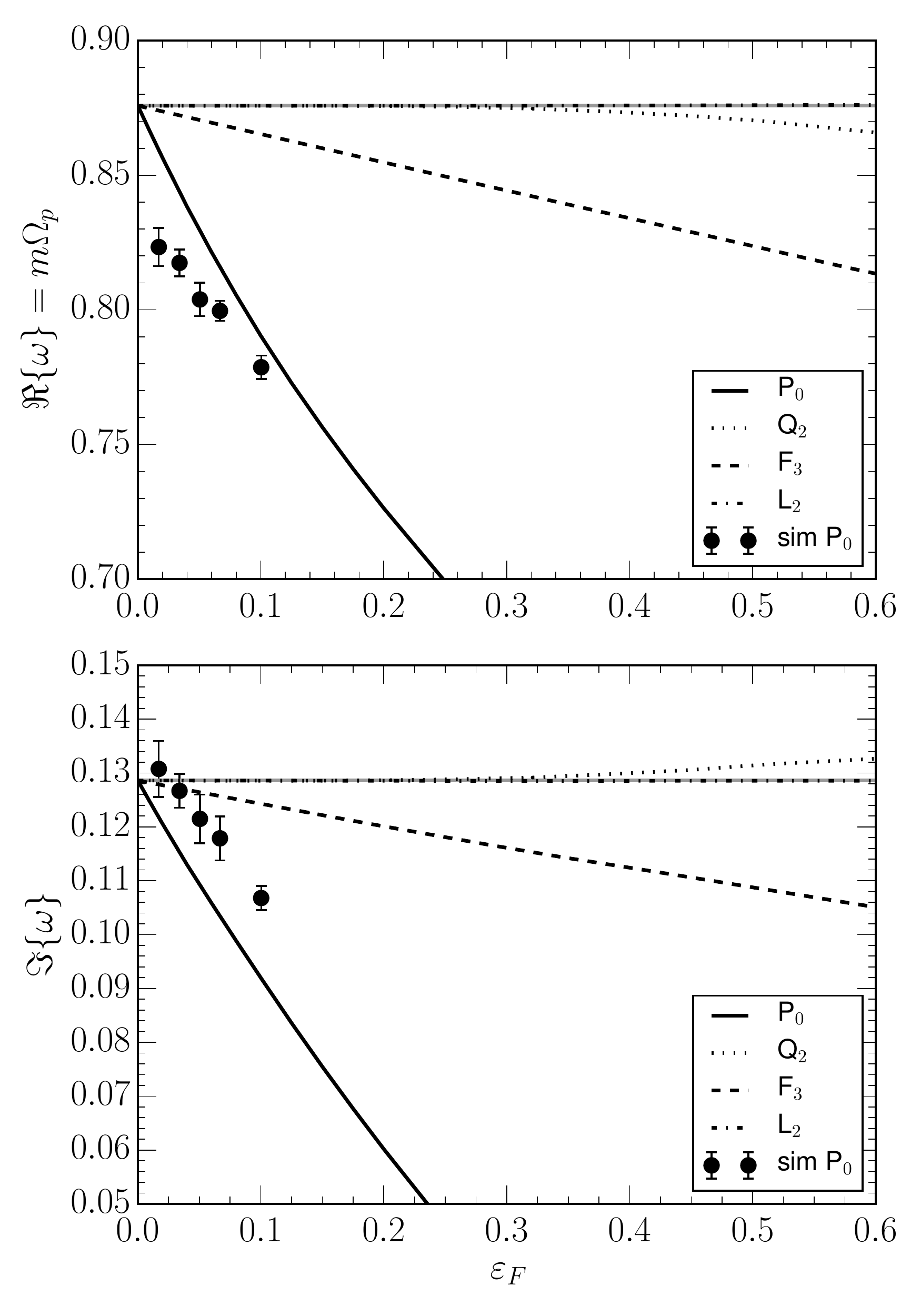}
\caption{Pattern speed $\Re\{ \omega \}=m\Omega_p$ (top panel) and
  growth rate $\Im\{ \omega \}$ (bottom panel) of the dominant $m=2$
  mode of the $q=6$, $n=4$ Mestel disc for different softening recipes
  (Plummer P$_0$, modified Kuz'min Q$_2$, Ferrers F$_3$, and modified
  Ferrers L$_2$). The softening lengths are scaled according to
  equations~\eqref{eq:eps:scaling} and \eqref{eq:eps:scale:force} to
  obtain a common maximum inter-particle force. The datapoints are
  derived from the $N$-body simulations by
  \citet{2001ApJ...546..176S}. The horizontal grey lines indicate the
  Newtonian mode frequency.
 \label{fig:Mestel_allsoft}}
\end{figure}

\subsubsection{Unsoftened gravity}

{\sc pyStab} retrieves the dominant mode of the $q=6$, $n=4$ Mestel
disc, along with a number of much slower growing modes. Since the
frequencies of these minor modes are sensitive to the choice of
numerical parameter values, they are most likely spurious
\citep{poly15}. Using unsoftened gravity, we find the dominant mode to
have a frequency $\omega=0.876+0.128\;\mathrm{i}$, which is in good
agreement with the values reported by \citet{1977ARA&A..15..437T,
  1997PhDT.........1R,poly15} and which were computed using different
mode analysis techniques and codes.

\subsubsection{Softened gravity}

In Fig. \ref{fig:Mestel_allsoft}, we show how the pattern speed (top
panel) and growth rate (bottom panel) of the dominant mode of this
base state change with increasing softening length (scaled according
to equations~\eqref{eq:eps:scaling} and \eqref{eq:eps:scale:force} to
a common maximum inter-particle force) using the softening recipes
listed in Table~\ref{tab:soft}.

Overplotted in this figure, we show the frequency estimates of
\citet{2001ApJ...546..176S} for this Mestel disc model, based on
$N$-body simulations with a particle-mesh code employing 2.5 million
particles, and a grid of 256 azimuthal and 200 radial nodes. The 5
simulations presented here all start from exactly the same initial
conditions but are evolved using different Plummer softening
lengths. The agreement with our linear mode analysis is not as good as
in the case of the isochrone disc. As reported by
\citet{2001ApJ...546..176S}, there is a $\sim 10$~\% scatter between
the measured frequencies of simulations with resampled initial
conditions at a constant particle number. This may be why the
simulation datapoints do not converge to the Newtonian linear-mode
result for zero softening length. Moreover, particle noise may have
negatively affected the frequency measurements. Still, the trend
followed by these simulations is in qualitative agreement with our
results:~both the pattern speed and the growth rate decrease with
increasing softening length.

As for the isochrone disc, softening methods with $a_0=0$ (like Q$_2$)
stay much closer to the Newtonian mode frequency than methods with
compact support but non-zero $a_0$ (like F$_3$) for a given value of
the softening length $\varepsilon$. Methods that combine compact
support with $a_0=0$ (like L$_2$) generally outperform the others.

Again, we refer the reader to section \ref{sec:dis} for a discussion
of how to correctly interpret this apparent success.

\section{Discussion} \label{sec:dis}

\subsection{Scaling to the same level of gravity bias}
\begin{figure*}
	\includegraphics[trim=0 12.5 0
          0,clip,width=0.49\textwidth]{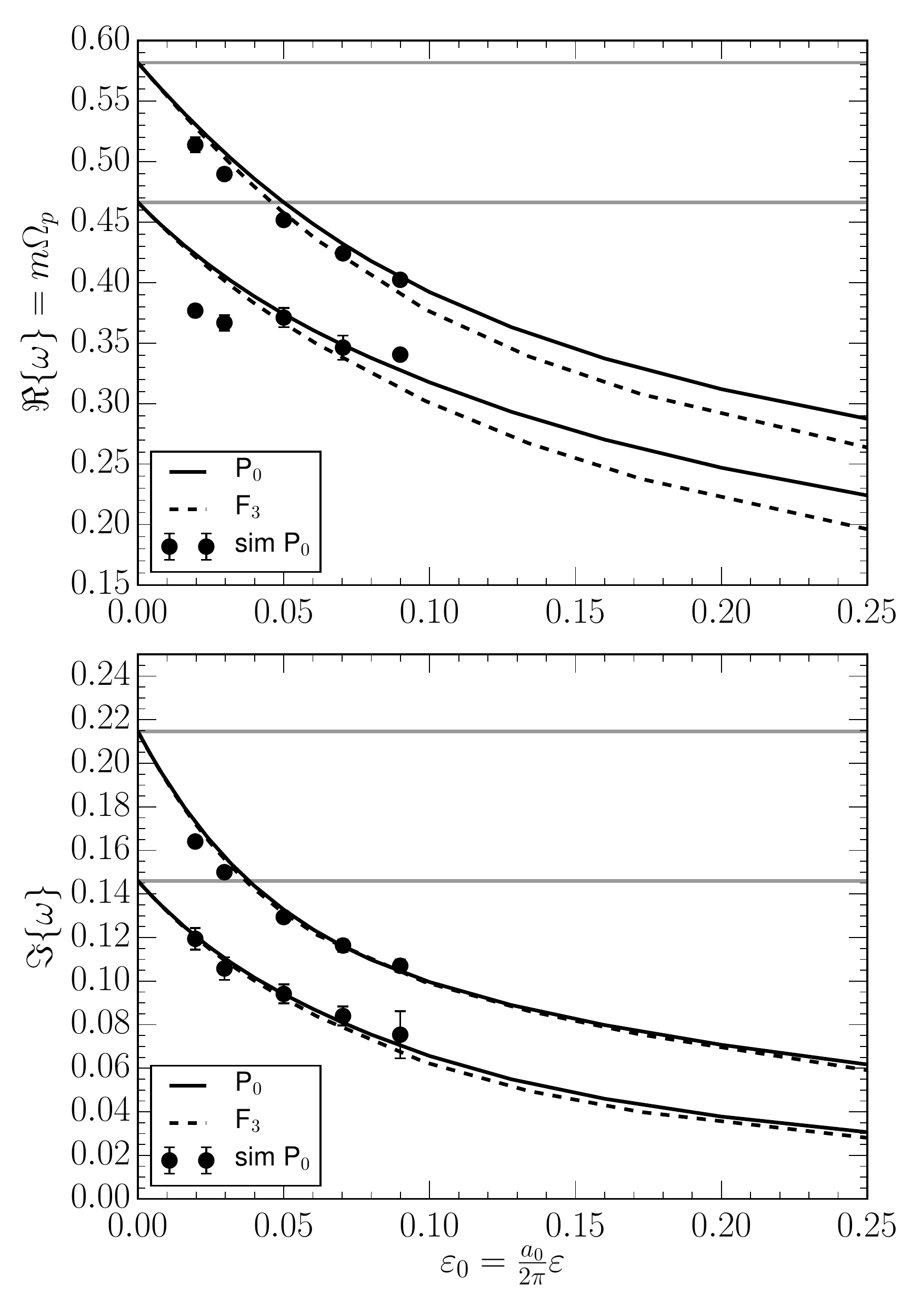}
	\includegraphics[trim=0 12.5 0
          0,clip,width=0.49\textwidth]{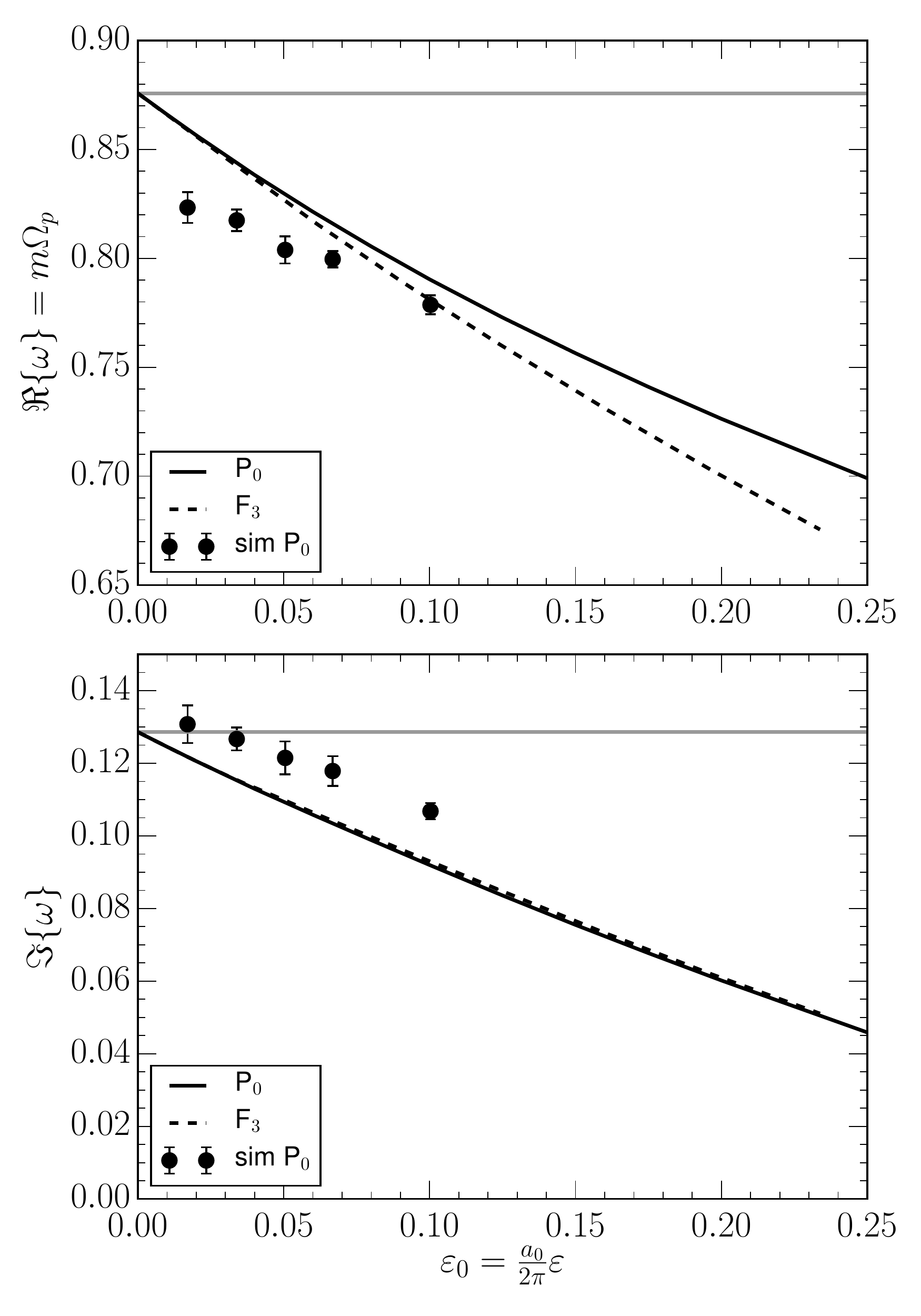}
	\caption{Pattern speed $\Re\{ \omega \}=m\Omega_p$ (top
          panels) and growth rate $\Im\{ \omega \}$ (bottom panels) of
          the dominant $m=2$ mode of the $m_K=12$ isochrone disc (left
          panels) and the $q=6$, $n=4$ Mestel disc (right panels) as a
          function of the scaled softening length $\varepsilon_0$,
          which allows for a direct comparison of the softening
          kernels with non-zero $a_0$ (i.e. P$_0$ and F$_3$) at the
          same level of gravity bias. The bullets are data taken from
          the $N$-body simulations using Plummer softening reported in
          \citet{1995ApJ...451..533E} and
          \citet{2001ApJ...546..176S}. The horizontal grey lines
          indicate the Newtonian mode frequencies.
	\label{fig:Isoa0}}
\end{figure*}

As mentioned in paragraph \ref{sec:soft:scale}, the softening length
and kernel are only defined up to a re-scaling and we advocate the
scale \beqn \varepsilon_0 = \frac{a_0}{2\pi}\eps \neqn to bring
methods with non-zero $a_0$ to a common gravity bias level.

In Fig. \ref{fig:Isoa0}, we show the retrieved frequencies of the
modes of the isochrone and Mestel discs as a function of
$\varepsilon_0$ for the two softening methods with non-zero gravity
bias parameter $a_0$ (i.e. P$_0$ and F$_3$). Clearly, the differences
between both softening methods, which are so striking in Figures
\ref{fig:Isochrone_allsoft} and \ref{fig:Mestel_allsoft}, now largely
disappear. At a given $\varepsilon_0$-value, all $a_0 \ne 0$ methods
perform almost equally well. Thus, it is always possible to re-scale
the softening length of one $a_0 \ne 0$ softening technique such that
it approximately matches the performance of another $a_0 \ne 0$
method. No exact matching is possible because of the higher order
terms in the expansion of the gravity bias.

\begin{figure*}
	\includegraphics[trim=0 12.5 0
          0,clip,width=0.49\textwidth]{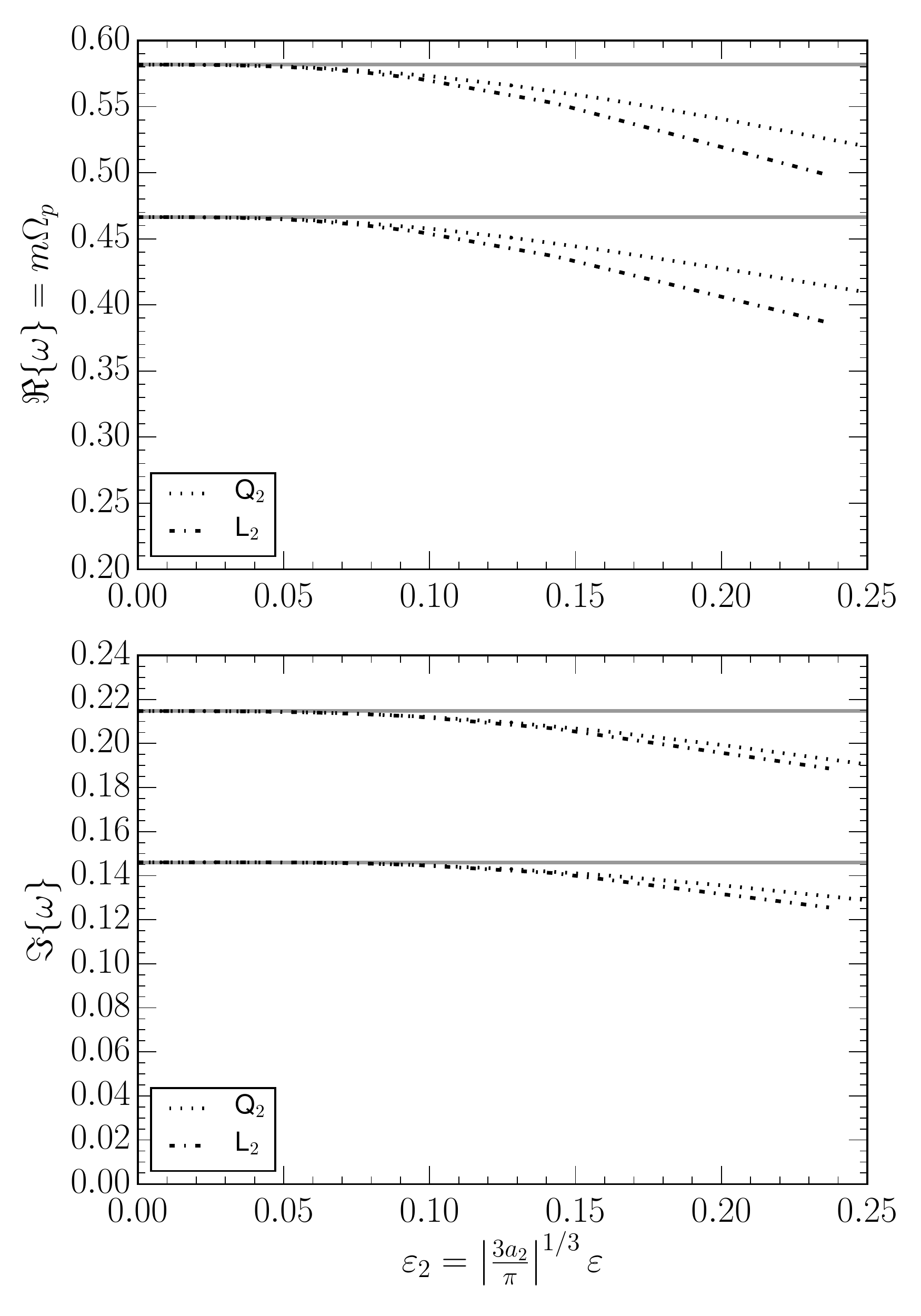}
	\includegraphics[trim=0 12.5 0
          0,clip,width=0.49\textwidth]{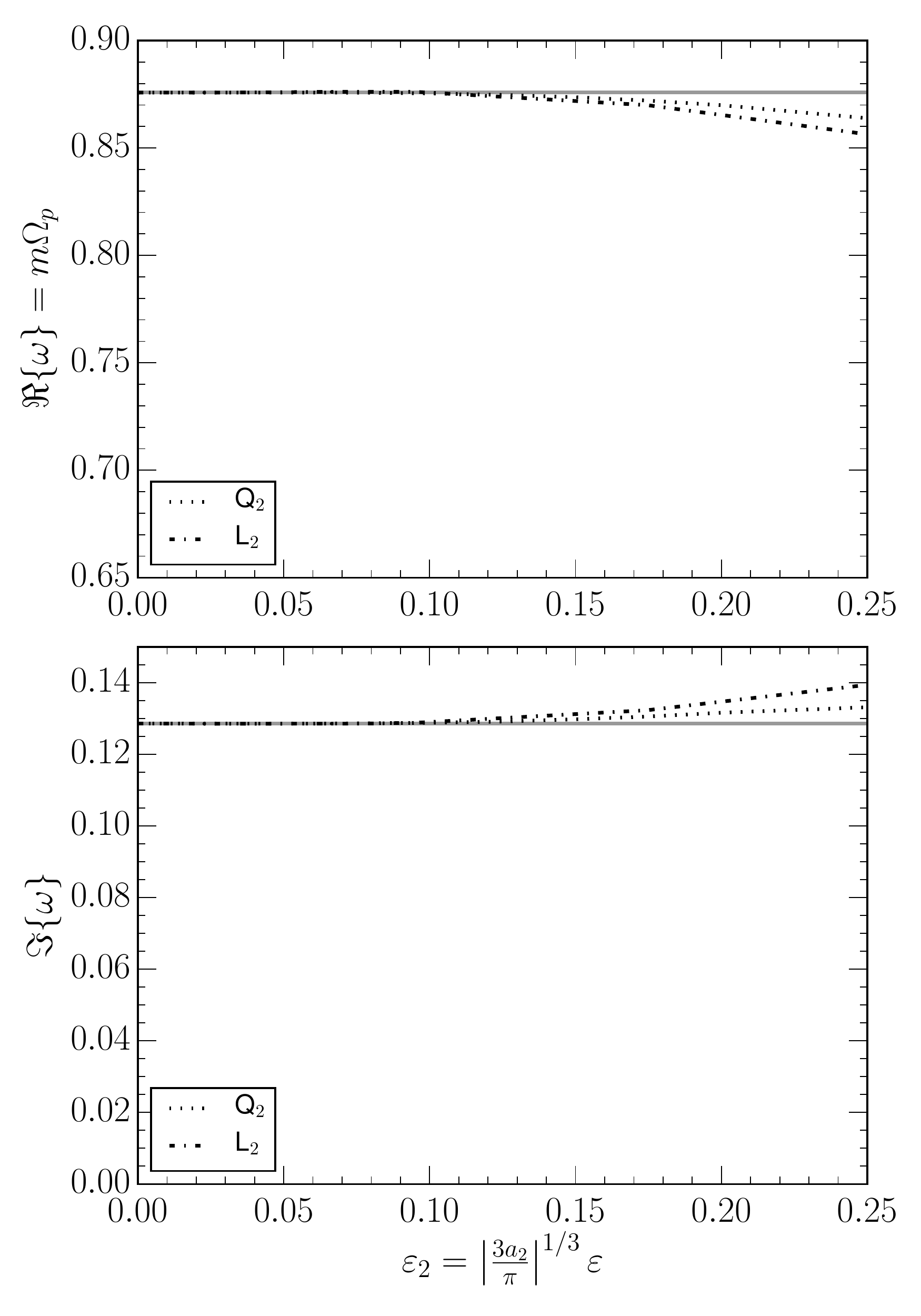}
	\caption{Pattern speed $\Re\{ \omega \}=m\Omega_p$ (top
          panels) and growth rate $\Im\{ \omega \}$ (bottom panels) of
          the dominant $m=2$ mode of the $m_K=12$ isochrone disc (left
          panels) and the $q=6$, $n=4$ Mestel disc (right panels) as a
          function of the scaled softening length $\varepsilon_2$,
          which allows for a direct comparison of the softening
          kernels with zero $a_0$ (i.e. Q$_2$ and L$_2$) at the same
          level of gravity bias. The horizontal grey lines indicate
          the Newtonian mode frequencies.
	\label{fig:Isoa2}}
\end{figure*}

As can be seen in Figure \ref{fig:Isoa2}, softening techniques with
$a_0=0$ but non-zero $a_2$ can be scaled to a common level of gravity
bias using the transformation (\ref{a2scal}), allowing for higher
order terms in the expansion (\ref{eq:bias:V:2D}) for the gravity
bias.

Based on these results, it seems fair to say that softening strategies
with $a_0=0$ generally yield more accurate (i.e. Newtonian-like) mode
frequencies than strategies with $a_0 \ne 0$ because the gravity bias
of the latter grows linearly with softening length $\varepsilon$ while
for the former it grows much more slowly, as $\varepsilon^3$. However,
within each of these classes of softening techniques, there is no
particular reason to favour one method over another provided they are
compared at (approximately) the same level of gravity bias.

\subsection{Physical interpretation}

We define the two-dimensional Fourier transform $\widehat{\pot}(k)$ of
the interparticle interaction potential $\pot(r)$ as
\begin{equation}
\pot(\vec{r}) = \frac{1}{(2\pi)^2}\int \widehat{\pot}(k) {\mathe}^{{\mathi}
  \vec{k}.\vec{r}} \mathrm{d}\vec{k}.
\end{equation}
Based on Poisson's equation, using separation of variables it is
straightforward to show that the radial part of the gravitational
response potential, which we denote here by $V_m(r)$, generated by an
$m$-armed spiral response density of the form
\begin{equation}
  \Sigma_m(r,\theta) = \Sigma_m(r) {\mathe}^{{\mathi} m \theta}
\end{equation}
can be retrieved from the relation
\begin{equation}
  {\mathcal H}_m \left\{ V_m(r) \right\} = - G
  \widehat{\pot}(k) {\mathcal H}_m\left\{ \Sigma_m(r) \right\}
\end{equation}
with ${\mathcal H}_m$ the Hankel transform of order $m$ \citep[see
  e.g.][]{bintrem}. The Hankel transform of order $m$ of a function
$f(r)$ is defined as
\begin{equation}
  {\mathcal H}_m\left\{ f \right\}(k) = \int_0^\infty f(r) J_m(kr) r
  \mathrm{d}r
\end{equation}
with $J_m(x)$ a Bessel function of the first kind.

Here, we will use the notation $\widehat{\pot}_N(k)$ for the Fourier
transform of the Newtonian interaction potential, with
\begin{equation}
  \widehat{\pot}_N(k) = \frac{2\pi }{k}.
\end{equation}
Likewise, for the interaction potentials listed in
Table~\ref{tab:soft} we find that:
\begin{align}
 \widehat{\pot}_{P_0}(k) &= 2\pi\frac{{\mathe}^{-k\varepsilon}}{ k} \label{kp0}
 \\ \widehat{\pot}_{Q_2}(k) &= 2\pi\left(1+k\varepsilon + \frac{1}{2}
 (k\varepsilon)^2\right)\frac{ {\mathe}^{-k\varepsilon}}{ k}. \label{kq2}
\end{align}
For the softening techniques with compact support, F$_3$ and L$_2$, no
simple analytical expression exists for the Fourier transform of their
interaction potentials but they can easily be obtained numerically.

\begin{figure}
\includegraphics[trim=5 15 10 10,clip,width=0.465\textwidth]{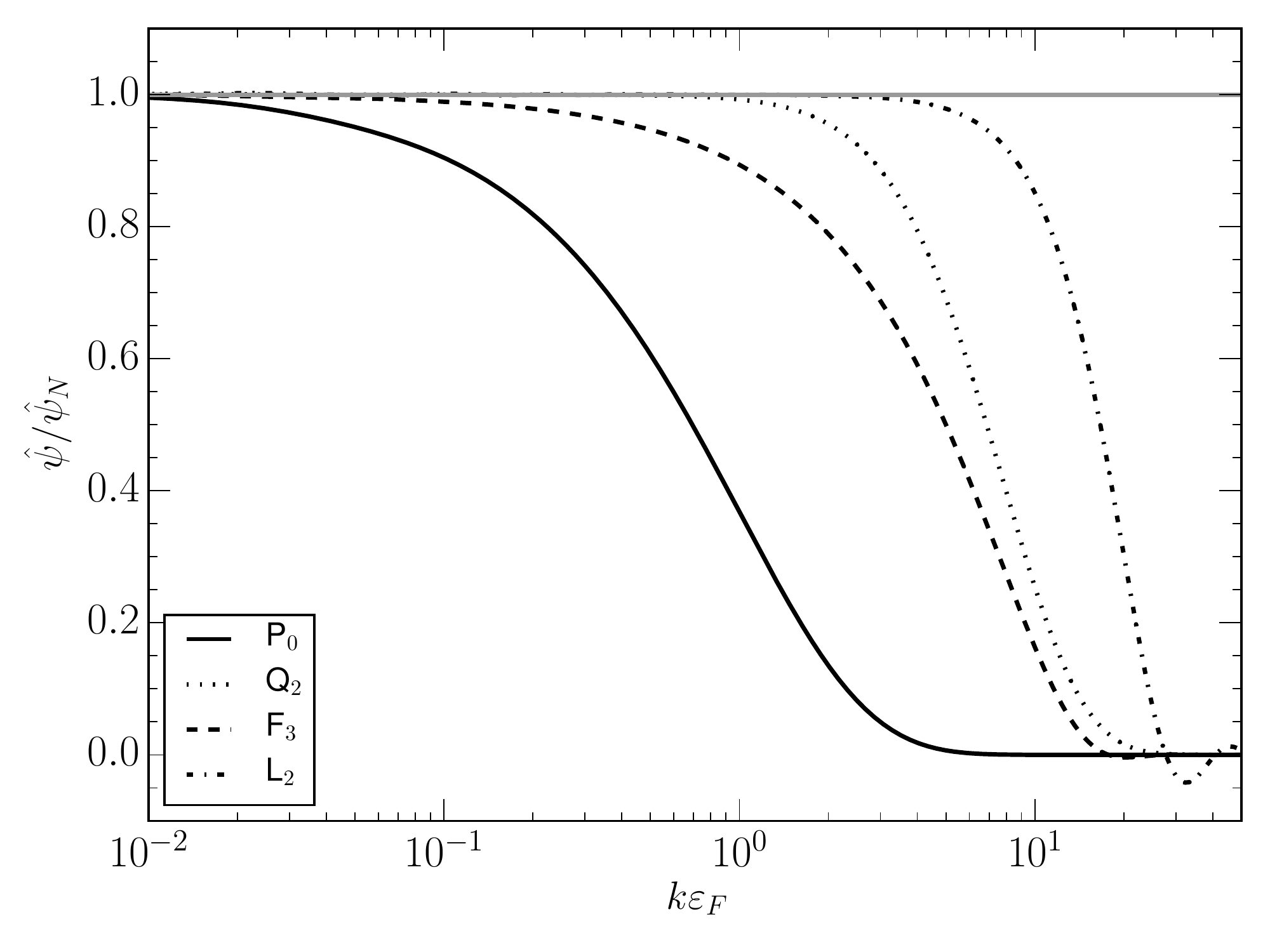}
\caption{The suppression factor $\widehat{\pot}/\widehat{\pot}_N$ as a
  function of the dimensionless wave number $k\varepsilon_F$, where
  the softening lengths are scaled according to
  equations~\eqref{eq:eps:scaling} and \eqref{eq:eps:scale:force} to
  obtain a common maximum inter-particle force.
 \label{fig:hank}}
\end{figure}

\begin{figure}
\includegraphics[trim=5 15 10 10,clip,width=0.475\textwidth]{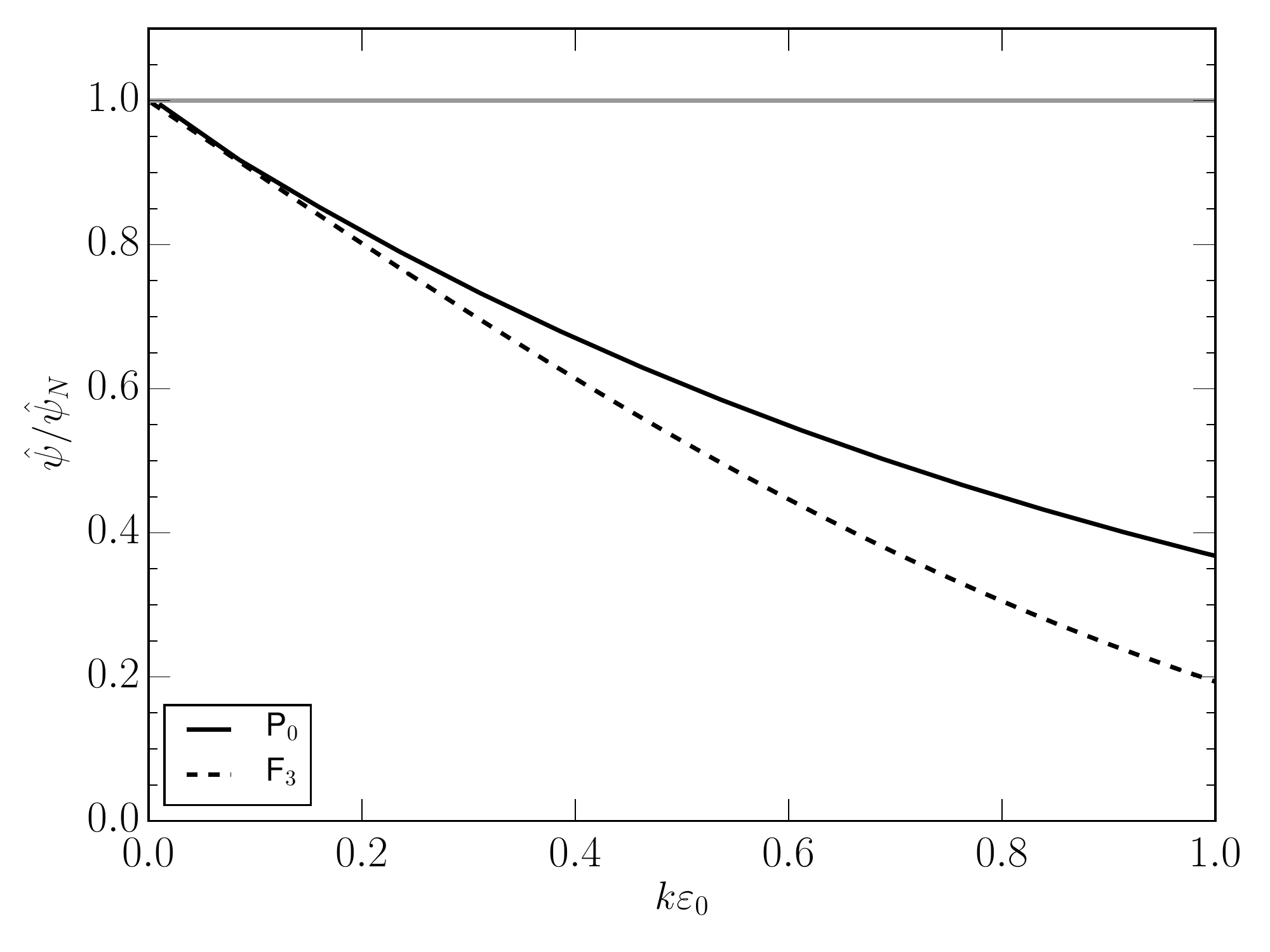}
\includegraphics[trim=5 15 10 10,clip,width=0.475\textwidth]{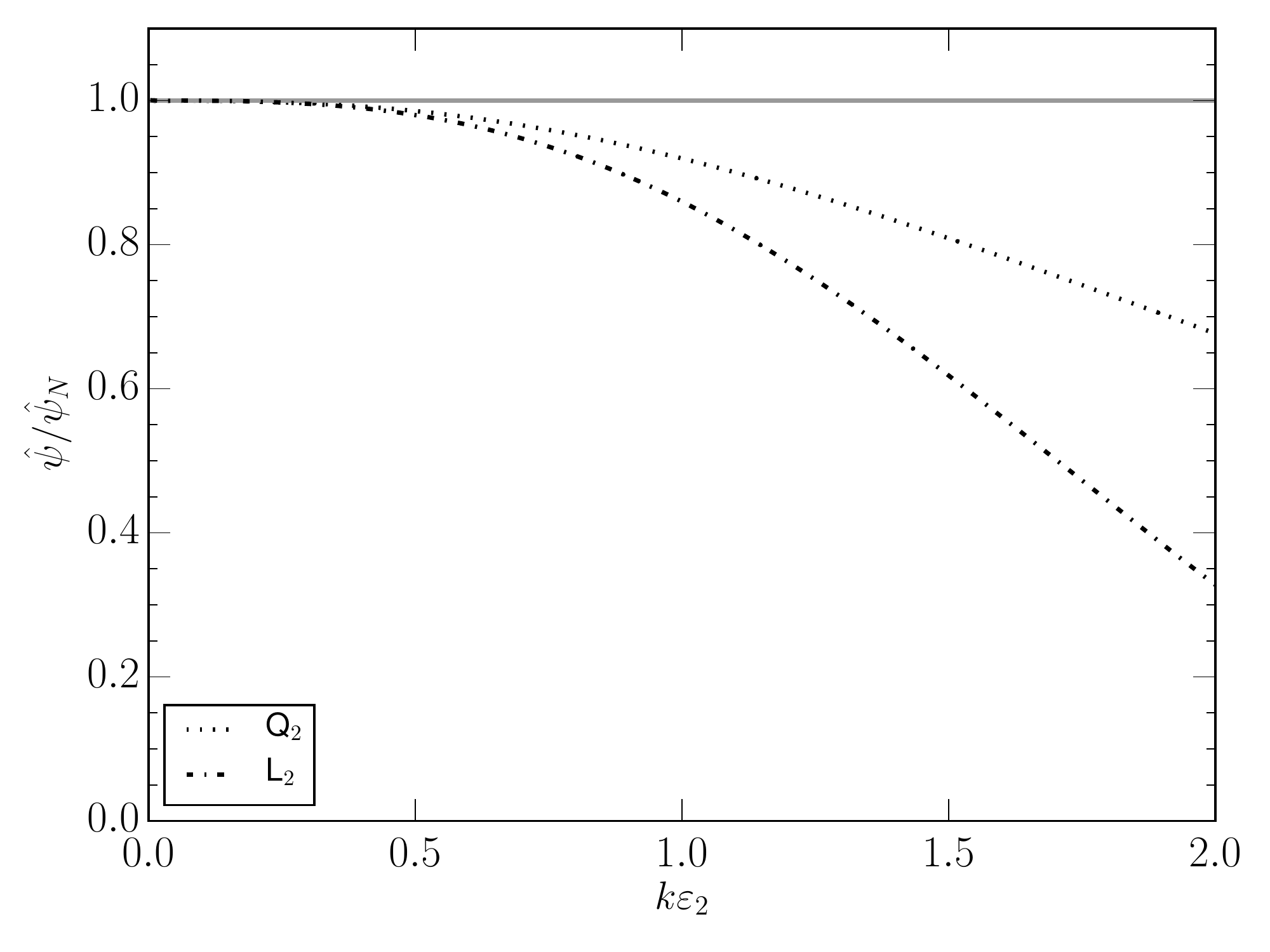}
\caption{The suppression factor $\widehat{\pot}/\widehat{\pot}_N$ as a
  function of the re-scaled dimensionless wave numbers
  $k\varepsilon_0$ (top panel), for the softening methods with $a_0
  \ne 0$, and $k\varepsilon_2$ (bottom panel), for the softening
  methods with $a_0 = 0$.
 \label{fig:hank02}}
\end{figure}

For a given response density $\Sigma_m$, the softened response
potential $V_m^\varepsilon$ and the unsoftened response potential
$V_m^0$ are connected as
\begin{equation}
  {\mathcal H}_m\left\{ V_m^\varepsilon \right\} =
  \frac{\widehat{\pot}}{\widehat{\pot}_N} {\mathcal H}_m\left\{ V_m^0
  \right\} 
\end{equation}
with $\varepsilon$ the softening length. We plot the $k$-dependent
suppression factor $\widehat{\pot}/\widehat{\pot}_N$ that links the
Fourier transforms of the softened and unsoftened response potentials
in Figure \ref{fig:hank}. The most striking consequence of
gravitational softening is the suppression of the small-scale,
i.e. large wavenumber $k$, structure in the Fourier transform of the
response potential.

In Appendix \ref{app1}, we show how this suppression factor is
connected to the gravity bias. More specifically, we prove that the
even coefficients in the series expansion of the suppression factor
around zero $k$ are directly proportional to the even coefficients in
the series expansion of the gravity bias around zero $\varepsilon$ (in
case the latter exist).
Clearly, desiging a softening kernel to
have vanishing bias coefficients is equivalent to designing an
interaction kernel whose suppression factor
$\widehat{\pot}/\widehat{\pot}_N$ is increasingly close to unity for
small wavenumbers $k$. 

However, just as it is not sensible to compare the various softening
strategies as we did in Figures \ref{fig:Isochrone_allsoft} and
\ref{fig:Mestel_allsoft}, it makes little sense to compare the
suppression factors as a function of $k \eps_F$. It is more meaningful
to compare the suppression factors at the same level of gravity bias,
i.e. as a function of $k \eps_0$ for the softening methods with $a_0
\ne 0$, and as a function of $k\varepsilon_2$ for the softening
methods with $a_0 = 0$. This comparison is shown in Figure
\ref{fig:hank02} and restates our previous conclusions. As a function
of the re-scaled wave number $k \varepsilon_0$, which places the P$_0$
and F$_3$ kernels on an equal gravity bias footing, the suppression
factors of these two softening techniques behave remarkably
similar. In fact, Plummer softening leads to less suppression for
large $k\varepsilon_0$ than F$_3$ softening. This agrees with Figure
\ref{fig:Isoa0} in which Plummer softening is shown to stay closer to
the correct, Newtonian result than F$_3$ softening at an equal level
of gravity bias. The Q$_2$ kernel, in turn, leads to less suppression
than the L$_2$ kernel and, as can be seen in Figure \ref{fig:Isoa2},
it also leads to slightly better frequency estimates.

This suppression of the response potential will likely lead to an
increased stability of the model galaxy. This expectation is borne out
by studying the stability of axially symmetric WKBJ waves under
Plummer softening, where a Toomre $Q$-value $Q<1$ now separates
growing from stationary waves \citep{1971Ap&SS..14...73M}. This
analysis has been extended to include general softening kernels by
\citet{1994A&A...286..799R,1997A&A...324..523R}. The physical
background of our results and the expected influence of softening on
$m=0$ WKBJ waves are, therefore, already well understood. Here, we
took this work further by studying general eigenmodes beyond the WKBJ
approximation and by going to $m=2$ patterns.

\section{Conclusions}

We use linear perturbation theory to investigate how different recipes
for gravitational softening, as employed in numerical $N$-body
simulations of razor-thin disc galaxies, affect predictions for the
properties of the latter's spiral eigenmodes. We specifically focus on
the frequencies, i.e. pattern speeds and growth rates, of two-armed
modes in the linear regime. We warn the reader that our approach does
not take into account the effects of approximate force evaluations
\citep{1986Natur.324..446B}, finite-$N$
\citep{2001MNRAS.324..273D,2012ApJ...751...44S,2015A&A...584A.129F},
stochasticity \citep{2009MNRAS.398.1279S}, implicit softening
contributed by the grid in particle-mesh codes
\citep{1994A&A...286..799R}, etc. whose respective magnitudes,
moreover, may depend on the amount of gravitational softening.

We have tested our linear mode analysis approach by comparing the
behaviour of the frequencies of the dominant $m=2$ modes of an
isochrone disc and of the Mestel disc as a function of Plummer
softening length with those found in the $N$-body simulations reported
by \citet{1995ApJ...451..533E} and
\citet{2001ApJ...546..176S}. Overall, we found reasonably good
agreement between linear theory and numerical simulations, also in the
softened regime.

We argue that the only meaningful way of comparing softening kernels
is to scale them to the same gravity bias level. In this paper, we
show how this scaling can be achieved, based on the results of
\citet{2001MNRAS.324..273D}. Thus, it is always possible to re-scale
the softening length of one softening technique such that it matches
the performance of another method with the same dependence of gravity
bias on softening length.

We have shown that softening methods with a vanishing lowest-order
term in the expansion of the gravity bias as a function of softening
length (in two dimensions, this is a linear term; in three dimensions,
this term is quadratic in the softening length) and whose gravity bias
therefore grows slowly with increasing softening length (e.g. the
Q$_2$ and L$_2$ methods discussed in this paper) provide more accurate
mode frequency estimates than methods with a non-zero lowest-order
term (e.g. the P$_0$ and F$_3$ methods). Softening methods with zero
lowest-order term compensate the sub-Newtonian forces deep inside the
kernel with super-Newtonian forces near $r\sim \eps$.


Kernels with compact support, in the sense that they yield exactly
Newtonian forces outside of the softening kernel, perhaps somewhat
counter-intuitively, do not necessarily provide more accurate
frequency estimates than kernels with infinite extent. For instance,
when compared at a common level of gravity bias, the Plummer kernel
(P$_0$) provides more accurate frequency estimates than the F$_3$
kernel. Likewise, the Q$_2$ kernel outperforms the L$_2$ kernel in
this regard.

The relative merit of a softening kernel can be judged from its
suppression of the small-scale, i.e. large wavenumber $k$, structure
in the Fourier transform of its response potential. The stronger this
suppression, measured at a given level of gravity bias, the more the
mode frequency estimates deviate from their Newtonian values.


As a guide to simulators, we provide an example of how a softening
technique, in this case Plummer softening, can be used as a basis for
developing new softening kernels whose gravity biases grow more slowly
with increasing softening length. These then provide much more
accurate estimates for mode frequencies than Plummer softening does.

Generally, the use of gravitational softening lowers the exponential
growth rate of spiral modes. Therefore, strongly softened $N$-body
simulations may risk ``losing'' some of these modes as their growth
rates are overtaken by that of e.g. swing amplified noise
\citep{1995A&AT....7..317R}.

\section*{Acknowledgements}

We wish to thank the organizers of the workshop ``The secular
evolution of self-gravitating systems over cosmic ages'' at the
Institut d'Astrophysique de Paris, May 24-27 2016, where this work was
initiated. We are grateful to Prof. Jerry Sellwood for his very
helpful insights into setting up and performing $N$-body simulations
and to Prof. Scott Tremaine for his helpful suggestions.  JBF
acknowledges support from Program number HST-HF2-51374 which was
provided by NASA through a grant from the Space Telescope Science
Institute, which is operated by the Association of Universities for
Research in Astronomy, Incorporated, under NASA contract
NAS5-26555. SDR acknowledges financial support from the European
Union's Horizon 2020 research and innovation programme under the Marie
Sk{\l}odowska-Curie grant agreement No 721463 to the SUNDIAL ITN
network. WD acknowledges support by STFC grant ST/N000757/1.

\bibliographystyle{mn2e}
\bibliography{manuscript}

\begin{thebibliography}{51}
\expandafter\ifx\csname natexlab\endcsname\relax\def\natexlab#1{#1}\fi

\bibitem[{{Barnes} \& {Hut}(1986)}]{1986Natur.324..446B}
{Barnes} J., {Hut} P., 1986, \nat, 324, 446

\bibitem[{{Barnes}(2012)}]{2012MNRAS.425.1104B}
{Barnes} J.~E., 2012, \mnras, 425, 1104

\bibitem[{{Binney} \& {Tremaine}(2008)}]{bintrem}
{Binney} J., {Tremaine} S., 2008, Galactic Dynamics, second edition. Princeton
  University Press

\bibitem[{{Colombi} {et~al}\mbox{.}(2015){Colombi}, {Sousbie}, {Peirani},
  {Plum}, \& {Suto}}]{2015MNRAS.450.3724C}
{Colombi} S., {Sousbie} T., {Peirani} S., {Plum} G., {Suto} Y., 2015, \mnras,
  450, 3724

\bibitem[{{De Rijcke} \& {Voulis}(2016)}]{dv16}
{De Rijcke} S., {Voulis} I., 2016, \mnras, 456, 2024

\bibitem[{{Dehnen}(2001)}]{2001MNRAS.324..273D}
{Dehnen} W., 2001, \mnras, 324, 273

\bibitem[{{D'Onghia}, {Vogelsberger} \& {Hernquist}(2013){D'Onghia},
  {Vogelsberger}, \& {Hernquist}}]{ovh13}
{D'Onghia} E., {Vogelsberger} M., {Hernquist} L., 2013, \apj, 766, 34

\bibitem[{{Dury} {et~al}\mbox{.}(2008){Dury}, {De Rijcke}, {Debattista}, \&
  {Dejonghe}}]{dury08}
{Dury} V., {De Rijcke} S., {Debattista} V.~P., {Dejonghe} H., 2008, \mnras,
  387, 2

\bibitem[{{Earn} \& {Sellwood}(1995)}]{1995ApJ...451..533E}
{Earn} D.~J.~D., {Sellwood} J.~A., 1995, \apj, 451, 533

\bibitem[{{Evans} \& {Read}(1998)}]{1998MNRAS.300..106E}
{Evans} N.~W., {Read} J.~C.~A., 1998, \mnras, 300, 106

\bibitem[{{Ferrers}(1877)}]{Ferrers}
{Ferrers} N.~M., 1877, Q.~J.~Pure Appl.~Math., 14, 1

\bibitem[{{Fouvry} {et~al}\mbox{.}(2015){Fouvry}, {Pichon}, {Magorrian}, \&
  {Chavanis}}]{2015A&A...584A.129F}
{Fouvry} J.~B., {Pichon} C., {Magorrian} J., {Chavanis} P.~H., 2015, \aap, 584,
  A129

\bibitem[{{Fridman} \& {Polyachenko}(1984)}]{polyfrid}
{Fridman} A.~M., {Polyachenko} V.~L., 1984, Physics of Gravitating Systems - I.
  Equilibrium and Stability. Springer-Verlag

\bibitem[{{Henon}(1959)}]{1959AnAp...22..126H}
{Henon} M., 1959, Annales d'Astrophysique, 22, 126

\bibitem[{{Hernquist} \& {Katz}(1989)}]{1989ApJS...70..419H}
{Hernquist} L., {Katz} N., 1989, \apjs, 70, 419

\bibitem[{{Hohl}(1971)}]{1971ApJ...168..343H}
{Hohl} F., 1971, \apj, 168, 343

\bibitem[{{Jalali}(2007)}]{jalali07}
{Jalali} M.~A., 2007, \apj, 669, 218

\bibitem[{{Jalali} \& {Hunter}(2005)}]{jalali05}
{Jalali} M.~A., {Hunter} C., 2005, \apj, 630, 804

\bibitem[{{Kalnajs}(1976)}]{1976ApJ...205..751K}
{Kalnajs} A.~J., 1976, \apj, 205, 751

\bibitem[{{Kalnajs}(1977)}]{kalnajs77}
{Kalnajs} A.~J., 1977, \apj, 212, 637

\bibitem[{{Kalnajs}(1999)}]{1999ASPC..165..325K}
{Kalnajs} A.~J., 1999, in Astronomical Society of the Pacific Conference
  Series, Vol. 165, The Third Stromlo Symposium: The Galactic Halo, {Gibson}
  B.~K., {Axelrod} R.~S., {Putman} M.~E., eds., p. 325

\bibitem[{{Kuz'min}(1956)}]{Kuzmin56}
{Kuz'min} G.~G., 1956, Astr. Zh., 33, 27

\bibitem[{{Mark}(1976)}]{1976ApJ...205..363M}
{Mark} J.~W.~K., 1976, \apj, 205, 363

\bibitem[{{Merritt}(1996)}]{1996AJ....111.2462M}
{Merritt} D., 1996, \aj, 111, 2462

\bibitem[{{Mestel}(1963)}]{1963MNRAS.126..553M}
{Mestel} L., 1963, \mnras, 126, 553

\bibitem[{{Miller}(1971)}]{1971Ap&SS..14...73M}
{Miller} R.~H., 1971, \apss, 14, 73

\bibitem[{{Monaghan}(1992)}]{1992ARA&A..30..543M}
{Monaghan} J.~J., 1992, \araa, 30, 543

\bibitem[{{Palmer}(1994)}]{palmer}
{Palmer} P.~L., 1994, Stability of Collisionless Stellar Systems. Kluwer
  Academic Publishers

\bibitem[{{Pichon} \& {Cannon}(1997)}]{1997MNRAS.291..616P}
{Pichon} C., {Cannon} R.~C., 1997, \mnras, 291, 616

\bibitem[{{Plummer}(1911)}]{1911MNRAS..71..460P}
{Plummer} H.~C., 1911, \mnras, 71, 460

\bibitem[{{Polyachenko} \& {Just}(2015)}]{poly15}
{Polyachenko} E.~V., {Just} A., 2015, \mnras, 446, 1203

\bibitem[{{Read}(1997)}]{1997PhDT.........1R}
{Read} J.~C.~A., 1997, PhD thesis, , University of Oxford

\bibitem[{{Romeo}(1992)}]{1992MNRAS.256..307R}
{Romeo} A.~B., 1992, \mnras, 256, 307

\bibitem[{{Romeo}(1994)}]{1994A&A...286..799R}
{Romeo} A.~B., 1994, \aap, 286, 799

\bibitem[{{Romeo}(1995)}]{1995A&AT....7..317R}
{Romeo} A.~B., 1995, Astronomical and Astrophysical Transactions, 7, 317

\bibitem[{{Romeo}(1997)}]{1997A&A...324..523R}
{Romeo} A.~B., 1997, \aap, 324, 523

\bibitem[{{Schaller} {et~al}\mbox{.}(2014){Schaller}, {Becker}, {Ruchayskiy},
  {Boyarsky}, \& {Shaposhnikov}}]{2014MNRAS.442.3073S}
{Schaller} M., {Becker} C., {Ruchayskiy} O., {Boyarsky} A., {Shaposhnikov} M.,
  2014, \mnras, 442, 3073

\bibitem[{{Sellwood}(1983)}]{1983JCoPh..50..337S}
{Sellwood} J.~A., 1983, Journal of Computational Physics, 50, 337

\bibitem[{{Sellwood}(2011)}]{2011MNRAS.410.1637S}
{Sellwood} J.~A., 2011, \mnras, 410, 1637

\bibitem[{{Sellwood}(2012)}]{2012ApJ...751...44S}
{Sellwood} J.~A., 2012, \apj, 751, 44

\bibitem[{{Sellwood} \& {Carlberg}(2014)}]{2014ApJ...785..137S}
{Sellwood} J.~A., {Carlberg} R.~G., 2014, \apj, 785, 137

\bibitem[{{Sellwood} \& {Debattista}(2009)}]{2009MNRAS.398.1279S}
{Sellwood} J.~A., {Debattista} V.~P., 2009, \mnras, 398, 1279

\bibitem[{{Sellwood} \& {Evans}(2001)}]{2001ApJ...546..176S}
{Sellwood} J.~A., {Evans} N.~W., 2001, \apj, 546, 176

\bibitem[{{Sellwood} \& {Kahn}(1991)}]{sellwood91}
{Sellwood} J.~A., {Kahn} F.~D., 1991, \mnras, 250, 278

\bibitem[{{Sellwood} \& {Lin}(1989)}]{sellwood89}
{Sellwood} J.~A., {Lin} D.~N.~C., 1989, \mnras, 240, 991

\bibitem[{{Springel}(2010)}]{2010MNRAS.401..791S}
{Springel} V., 2010, \mnras, 401, 791

\bibitem[{{Toomre}(1977)}]{1977ARA&A..15..437T}
{Toomre} A., 1977, \araa, 15, 437

\bibitem[{{Toomre}(1981)}]{toomre81}
{Toomre} A., 1981, in Structure and Evolution of Normal Galaxies, {Fall} S.~M.,
  {Lynden-Bell} D., eds., pp. 111--136

\bibitem[{{Vauterin} \& {Dejonghe}(1996)}]{b9}
{Vauterin} P., {Dejonghe} H., 1996, \aap, 313, 465

\bibitem[{{Yoshikawa}, {Yoshida} \& {Umemura}(2013){Yoshikawa}, {Yoshida}, \&
  {Umemura}}]{2013ApJ...762..116Y}
{Yoshikawa} K., {Yoshida} N., {Umemura} M., 2013, \apj, 762, 116

\bibitem[{{Zang}(1976)}]{1976PhDT........26Z}
{Zang} T.~A., 1976, PhD thesis, Massachusetts Institute of Technology,
  Cambrigde, MA

\end{thebibliography}

\appendix

\section{The gravity bias and the suppression factor} \label{app1}
\subsection{Gravity bias}
Analogous to the three-dimensional case discussed in
\citet{2001MNRAS.324..273D}, in two dimensions, the expectation value
of the gravitational potential is
\begin{align}
	\left\langle\hat{V}(\vec{r})\right\rangle
  	&= -G\int \Sigma(\vec{r}')\,\psi(|\vec{r}-\vec{r}'|)\, \d^2\vec{r}', 
\end{align}
with $\Sigma$ the surface density that causes the gravitational
potential $V$ through the inter-particle interaction potential $-\psi$.
The integral covers the whole
surface of the galaxy. The gravity bias can be obtained by rewriting
the interaction potential as
\begin{equation}
  \psi(r) = 
  \frac{1}{\varepsilon}  \phi \left( \frac{r}{\varepsilon} \right)
 = \frac{1}{r} - \frac{1}{\varepsilon} \left[\frac{\varepsilon}{r} -
 \phi\left(\frac{r}{\varepsilon}\right)  \right], 
\end{equation}
which leads to
\begin{align}
  \mathrm{bias}_{\vec{r}}(\hat{V}) &\equiv \left\langle\hat{V}(\vec{r})\right\rangle-V(\vec{r}) \nonumber \\
  &  = \varepsilon G \int \Sigma(\vec{r}-\varepsilon \vec{u})\left[\frac{1}{u}-\phi(u) \right] \d^2\vec{u},
\end{align}
with $\varepsilon \vec{u} = \vec{r}-\vec{r}'$. We replace the surface
density by its Taylor expansion around the position $\vec{r}$,
\begin{equation}
  \Sigma(\vec{r}-\varepsilon \vec{u}) = \sum_{n = 0}^\infty
  \frac{(-\varepsilon)^n}{n!} \left( \vec{u}.\vec{\nabla} \right)^n
  \Sigma(\vec{r}).
  \end{equation}
such that
\begin{align}
	\mathrm{bias}_{\vec{r}}(V) &= \eps G \sum_{n=0}^\infty \frac{(-\eps)^{n}}{n!}
	\int \left[\frac{1}{u}-\phi(u)\right] (\vec{u}\cdot\vec{\nabla})^n \Sigma(\vec{r}) \,\d^2\vec{u}.
\end{align}
If we perform the integration in polar coordinates $(u,\theta)$ and replace $(\vec{u}\cdot\vec{\nabla})^n=u^n(\cos\theta\,\nabla_{\!x}+\sin\theta\,\nabla_{\!y})^n$ by its binomial expansion, we have
\begin{align}
	\mathrm{bias}_{\vec{r}}(V) &= \eps\,G \sum_{n=0}^\infty \frac{(-\eps)^{n}}{n!}
	\int_0^\infty u^{n}\left[1-u\phi(u)\right]\,\d u
	\nonumber \\ &\phantom{=} \times
	\label{eq:bias:intermediate:1}
	\sum_{l=0}^{n} \binom{n}{l} 
	\nabla_{\!x}^{l}\nabla_{\!y}^{n-l} \Sigma(\vec{r})
	\int_0^{2\pi} \cos^{l}\theta\sin^{n-l}\theta\,\d\theta.
\end{align}
The $\theta$ integral vanishes for odd $n$ or $l$, while for even $n$ and $l$
\begin{align}
	\label{eq:cos^lsin^n-l}
	\int_0^{2\pi} \cos^{l}\theta\sin^{n-l}\theta\,\d\theta
	&= \frac{2\pi n!}{2^n([n/2]!)^2} \binom{n/2}{l/2}\binom{n}{l}^{-1},
\end{align}
such that we obtain (using $\Delta\equiv\vec{\nabla}^2$)
\begin{align}
	\mathrm{bias}_{\vec{r}}(V) &= 
	\sum_{k=0}^\infty \eps^{2k+1} a_{2k}\,G\,
	\Delta^{k}\Sigma(\vec{r})
\end{align}
with coefficients $a_n$ as given by equation~\eqref{an}.
  
\subsection{Relation to the reduction factor}
Because $f(\vec{r})\equiv r^{-1}-\psi(r)$ is an isotropic function in $\vec{r}$ space, its moments
\begin{align}
	\mu_{l,m} \equiv \int x^l\,y^m\,f(\vec{r})\,d^2\!\vec{r}
\end{align}
vanish for odd $l$ or $m$. For even $l$ and $m$,
\begin{align}
	\mu_{l,m}
	&= \eps^{n+1}\,a_n n! \binom{n/2}{l/2} \binom{n}{l}^{-1},
\end{align}
where $n=l+m$ and we have used relation~\eqref{eq:cos^lsin^n-l}. Let
\begin{equation}
	\label{eq:F(k)}
	F(\vec{k}) \equiv \int f(\vec{r})\, \mathrm{e}^{-i\vec{k}\cdot\vec{r}}\,\d^2\!\vec{r}
\end{equation}
be the two-dimensional Fourier transform of $f(\vec{r})$. Because $f(\vec{r})$ is isotropic and real-valued, then so is $F(\vec{k})=F(k)$. From the equation~\eqref{eq:F(k)}
\begin{align}
	\left.\frac{\partial^{l+m}F}{\partial k_x^l\,\partial k_y^m}\right|_{\vec{k}=0}
	&=(-i)^{l+m} \mu_{l,m},
\end{align}
and hence the Taylor expansion of $F(\vec{k})$
\begin{align}
	F(\vec{k}) &= \sum_{l,m=0}^\infty \frac{k_x^l\,k_y^m}{l!\,m!} (-i)^{l+m}\mu_{l,m}
	\label{eq:FT:Taylor}
	= \eps \sum_{\nu=0}^\infty (-1)^{\nu} \, a_{2\nu}\, |\eps\vec{k}|^{2\nu}.
\end{align}
The reduction factor is related to $F(\vec{k})$ via
\begin{align}
	R(k)\equiv \widehat{\psi}(k)/\widehat{\psi}_N(k)
	&=1-(2\pi)^{-1}k F(k).
\end{align}
Thus, the coefficients $c_n$ of the Taylor series
\begin{align}
	R(k) = 1 - \sum_{n=0}^\infty c_n\,(\eps k)^{n+1}
\end{align}
are given by
\begin{align}
	\label{eq:cn:an}
	c_n = (-1)^{n/2} a_n/2\pi
\end{align}
for even $n$ and if $a_n$ is finite. For odd $n$ and/or infinite $a_n$, the situation is more complicated. If the kernel $\psi$ has compact support or $\psi\sim r^{-1}$ exponentially fast as $r\to\infty$, then all the coefficients $a_n$ are finite, the Taylor series~\eqref{eq:FT:Taylor} of $F(\vec{k})$ converges, and relation~\eqref{eq:cn:an} holds for all $n$, i.e.\ $R(k)$ is a function of $k^2$ only. This is the situation for the $F_3$ and $L_2$ kernels.

If the kernel does not satisfy the above conditions, but $f(r)\sim r^{-p}$ at $r\to\infty$, then $a_n=\infty$ for $n\ge p-2$ and the Taylor series~\eqref{eq:FT:Taylor} does not converge. However, the series of the non-divergent terms is still useful, only the remainder grows faster than $\eps^{p-2}$.
A typical example is Plummer softening, for which $a_0=2\pi$, while $a_{n>0}=\infty$ and (see equation~\ref{kp0})
\begin{equation}
	\label{eq:FT:Plummer}
	F(\vec{k}) = 2\pi k^{-1} (1-\mathrm{e}^{-\eps k})
	= 2\pi \left[\eps - \tfrac12\eps^2k 
	\ldots \right],
\end{equation}
i.e.\ $c_0=a_0/2\pi$ as per relation~\eqref{eq:cn:an}, but $c_1=-\tfrac12\neq0$. The problem is that the \emph{two-dimensional} Fourier transform~\eqref{eq:FT:Plummer} is not smooth at the origin, but has discontinuous gradient (such that its Taylor series fails), while the \emph{one-dimensional} function $F(k)$ and hence the reduction factor $R(k)$ are well behaved for all $k\ge0$. Thus, a divergent $a_n$ indicates a non-vanishing $c_{n-1}$ and, conversely, a $c_{n-1}\neq0$ for even $n$ implies $a_{n}=\infty$.

\label{lastpage}

\end{document}